\def\verARXIV{1}

\def\ver{1}

\ifx\ver\verARXIV

\documentclass[11pt,twoside,a4paper,note,final]{cms-tdr}
\usepackage{natbib}
\def\svnVersion{142936M}

\begin{document}

\hyphenation{had-ron-i-za-tion}
\hyphenation{cal-or-i-me-ter}
\hyphenation{de-vices}

\fi

\RCS$Revision: 142904 $
\RCS$HeadURL: svn+ssh://pjanot@svn.cern.ch/reps/tdr2/notes/pjanot_002/trunk/pjanot_002.tex $
\RCS$Id: pjanot_002.tex 142904 2012-08-08 08:01:51Z pjanot $

\newlength\cmsFigWidth
\ifthenelse{\boolean{cms@external}}{\setlength\cmsFigWidth{0.85\columnwidth}}{\setlength\cmsFigWidth{0.4\textwidth}}
\ifthenelse{\boolean{cms@external}}{\providecommand{\cmsLeft}{top}}{\providecommand{\cmsLeft}{left}}
\ifthenelse{\boolean{cms@external}}{\providecommand{\cmsRight}{bottom}}{\providecommand{\cmsRight}{right}}
\cmsNoteHeader{2012/003} 
\title{Prospective Studies for LEP3\\ with the CMS Detector\\ 
{\small \it submitted to the European Strategy Preparatory Group}}

\address[cern]{CERN, Geneva}
\address[mit]{Massachusetts Institute of Technology}
\address[padova]{INFN, Sezione di Padova}
\author[padova]{Patrizia Azzi}
\author[cern]{Colin Bernet}
\author[cern]{Cristina Botta}
\author[mit]{Guillelmo Gomez-Ceballos}
\author[cern]{Patrick Janot}
\author[mit]{\footnote[0]{Contact: Patrick.Janot@cern.ch} Markus Klute}
\author[cern]{Piergiulio Lenzi}
\author[cern]{Luca Malgeri}
\author[mit]{Marco Zanetti}

\date{\today}

\abstract{
On July 4, 2012, the discovery of a new boson, with mass around 125\,GeV/$c^2$ and with properties compatible with those of a standard-model Higgs boson, was announced at CERN.~\ In this context, a high-luminosity electron-positron collider ring, operating in the LHC tunnel at a centre-of-mass energy of 240 GeV and called LEP3, becomes an attractive opportunity both from financial and scientific point of views. The performance and the suitability of the CMS detector are evaluated, with emphasis on an accurate measurement of the Higgs boson properties. The precision expected for the Higgs boson couplings is found to be substantially better than that predicted by Linear Collider studies. 
}

\hypersetup{%
pdfauthor={Patrick Janot},%
pdftitle={Prospective Studies for LEP3 with the CMS Detector},%
pdfsubject={CMS},%
pdfkeywords={CMS, LEP3, Higgs, physics, software, upgrade}}


\maketitle

 
\section{Introduction}

With the analysis of 10\,${\rm fb}^{-1}$ of proton-proton collisions collected at centre-of-mass energies of 7 and 8\,TeV, the CMS and ATLAS collaborations have announced on July 4, 2012, the discovery~\cite{CMSHiggs,ATLASHiggs} of a new boson with a mass in the vicinity of $125$\,GeV/$c^2$. This particle decays into $\gamma \gamma$, ZZ and WW with rates tantalizingly close to those predicted for the standard-model (SM) Higgs boson. Other decay channels like ${\rm b \bar b}$ or $\tau^+\tau^-$ have been studied by CMS with their entire data sample, but the statistical uncertainties are still too large to draw any solid conclusions for these channels. The Tevatron also reported a broad excess in the mass region 120-135\,GeV/$c^2$ with the ${\rm b \bar b}$  channel~\cite{tevatron}.  Should the additional LHC data expected in 2012 confirm both the nature of the discovery and the values of the branching fractions, the next steps would naturally be detailed studies and precision measurements of this unique particle.

Some measurements are being performed by the LHC already in 2012, and will progressively improve with increasing centre-of-mass energy and integrated luminosity. With 300\,${\rm fb}^{-1}$ recorded at 13\,TeV, expected to happen by the end of the decade, SM-like couplings can be measured with a moderate precision~\cite{CMSESG}, typically 5 to 15\% for the couplings to gauge bosons, quarks and leptons in simplified models~\cite{zanetti}. Most of these measurements are expected to be limited by systematic uncertainties inherent to hadron collisions, so that additional luminosity might not help much in this respect. Ultimate precision measurements of the SM Higgs boson mass and width would have to wait for the advent of a technologically challenging muon collider~\cite{MuonCollider} at $\sqrt{s}=125$\,GeV. Meanwhile, the couplings of the Higgs boson to all other particles can be measured to a few per cent precision in ${\rm e}^+{\rm e}^-$ collisions, as demonstrated in extensive studies performed both at LEP and for a possible future high-energy linear collider, and summarized in Ref.~\cite{ilc} for the ILC and CLIC studies. 

The conclusion of these studies are that an  ${\rm e}^+{\rm e}^-$ linear collider is an excellent candidate for being the next collider at the energy frontier. More recently, however, a high-luminosity ${\rm e}^+{\rm e}^-$ collider operating in the LHC tunnel at $\sqrt{s}=240$\,GeV, hereafter conveniently called LEP3, was proposed~\cite{LEP3Proposal} to study the Higgs boson properties in detail. The major argument in favour of this proposal is the cost, substantially smaller than that of a linear collider: the tunnel and the infrastructure exist; two well-understood detectors are already in place; only a small fraction of the cavities needed for ILC would suffice to compensate for the 7\,GeV lost per turn by synchrotron radiation; and the optic design is a spin-off of the studies of a lepton-proton collider (LHeC) in the LHC tunnel. During a recent workshop~\cite{EUCARD} that took place at CERN on June 18, 2012, no unsurmountable obstacle has been identified for the realization of this high-luminosity collider at the beginning of the next decade~\cite{ESPP}, and the physics case was confirmed.

It is the purpose of this note {\it (i)} to summarize the LEP3 physics case and some of the LEP3 unique characteristics (Section~\ref{sec:LEP3}); {\it (ii)} to verify the capabilities of the CMS detector to proceed with precision measurements of the Higgs boson properties (Section~\ref{sec:CMS}); and {\it (iii)} to give a comparison with the linear collider physics potential (Section~\ref{sec:LC}).

\section{The LEP3 Physics case}
\label{sec:LEP3}

\subsection{The Physics Landscape in 2017}

\begin{figure*}[hbtp]
\begin{center} 
\begin{picture}(250,200)
\put(-100,18){\includegraphics[width=0.43\textwidth]{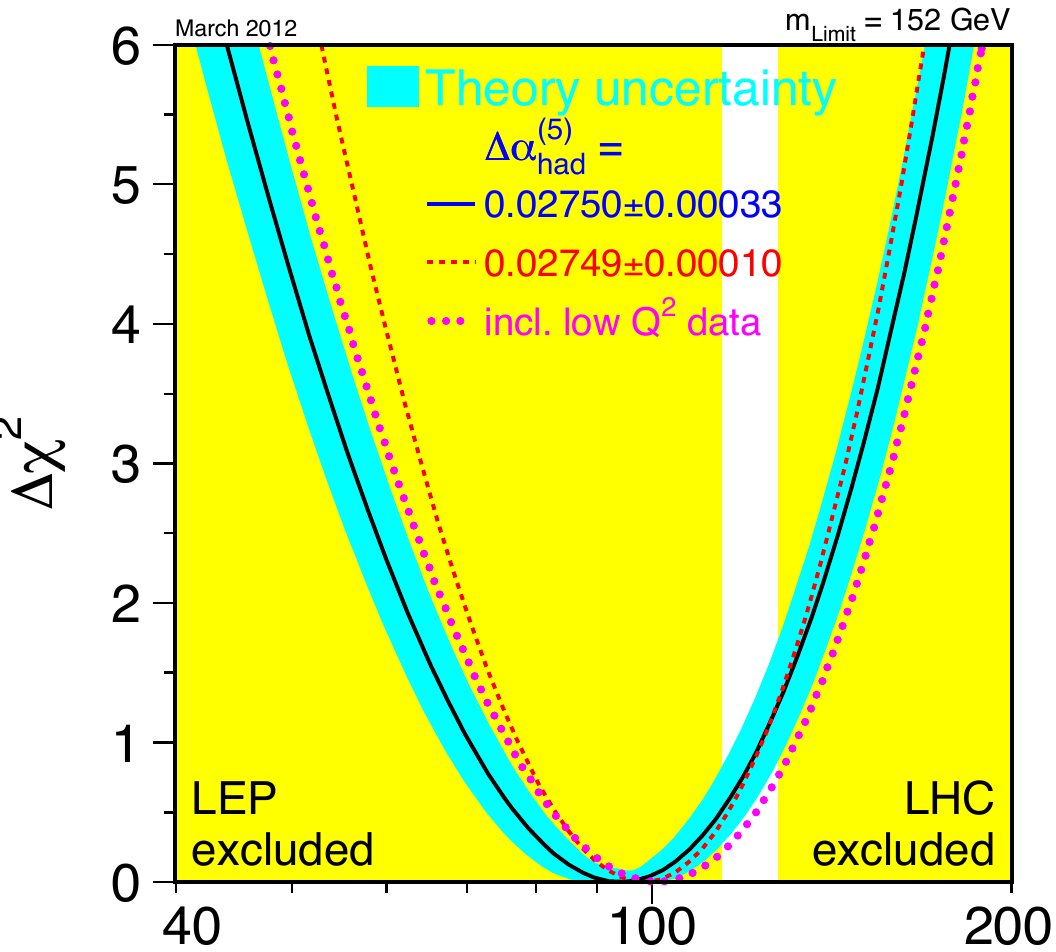}}
\put(110,0){\includegraphics[width=0.45\textwidth]{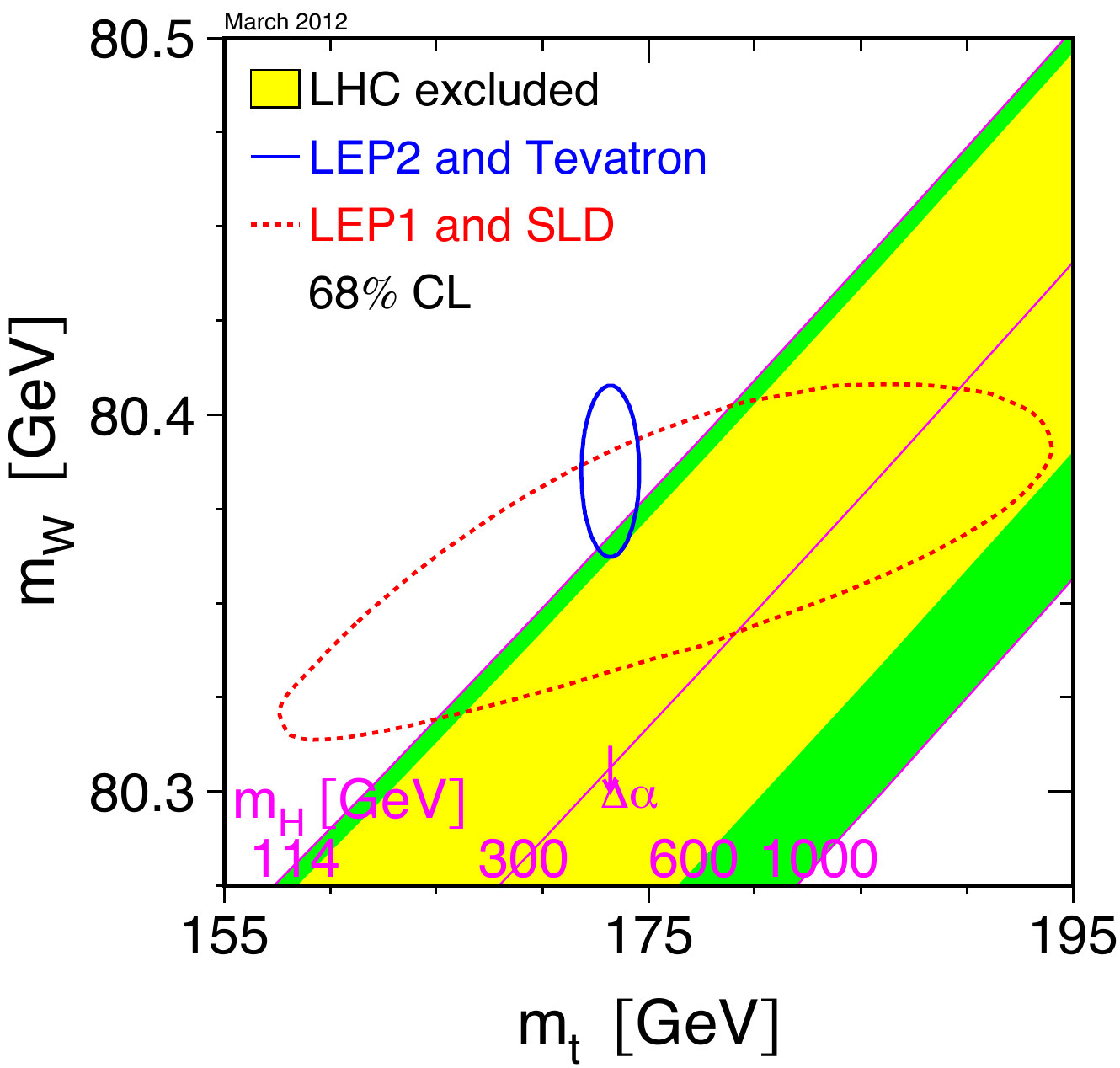}}
\end{picture}
 \caption{\small An executive overview of the physics landscape in March 2012, just prior to the Higgs boson discovery announcement. Details are given in the text.}
  \label{fig:mtmw}
\end{center}
\end{figure*}

The data analysed in the past 25 years, from LEP, SLC, TeVatron and LHC, allow the SM to be tested at the per mil level. The W and Z gauge bosons and the top quark were shown to be produced with the expected inclusive and differential cross sections. Even more importantly, these particles happen to have the expected masses: 
\begin{itemize}
\item the W mass is measured by LEP2 and the TeVatron with a combined precision of 15\,MeV/$c^2$, precisely where the Higgs mechanism predicts it to be from the Z mass and couplings measured at LEP1 and SLC: $m_{\rm W} = m_{\rm Z} \cos \theta_{\rm W}$ at lowest order, with known corrections from top-quark and Higgs-boson loops;
\item the top quark mass is measured at the TeVatron and the LHC with a precision of 1\,GeV/$c^2$, precisely at the value inferred by the magnitude of top-quark loop effects on all Z pole measurements;
\item the Higgs boson mass is now measured by CMS with a precision of 600\,MeV/$c^2$, and is in agreement with the value inferred from the Higgs-boson loop effects on all Z pole measurements associated to the W and top mass measurements.
\end{itemize}
The situation was summarized in Moriond 2012, {\it i.e.}, a couple months before the discovery announcement, with the graphs shown in Fig.~\ref{fig:mtmw}. On the left, the $\chi^2$ of all precision measurement combination is shown as a function of the Higgs boson mass in the context of the SM. This combination leads to the following prediction
\begin{equation}
m_{\rm H} = 94^{+29}_{-24}\, {\rm GeV}/c^2.
\end{equation}
Together with the result of the direct searches at LEP1 and LEP2, this prediction allows the SM Higgs boson mass range to be restricted to $\left[ 114.4, 152 \right]$\,GeV/$c^2$ at the 95\% confidence level. The direct searches conducted at the LHC in 2011 further restricts this allowed mass range to $\left[ 122, 127 \right]$\,GeV/$c^2$. The grand summary is displayed in the well known graph on the right of Fig.~\ref{fig:mtmw}, in which the direct top and W mass measurements (blue ellipse) are compared to the corresponding indirect measurements (red blob). The Higgs-boson-mass-dependent SM prediction for $122 <m_{\rm H} < 127$\,GeV/$c^2$, represented by the sole narrow green band that crosses both the blue and the red ellipses, is superimposed.

\begin{figure*}[hbtp]
\begin{center} 
\includegraphics[width=0.58\textwidth]{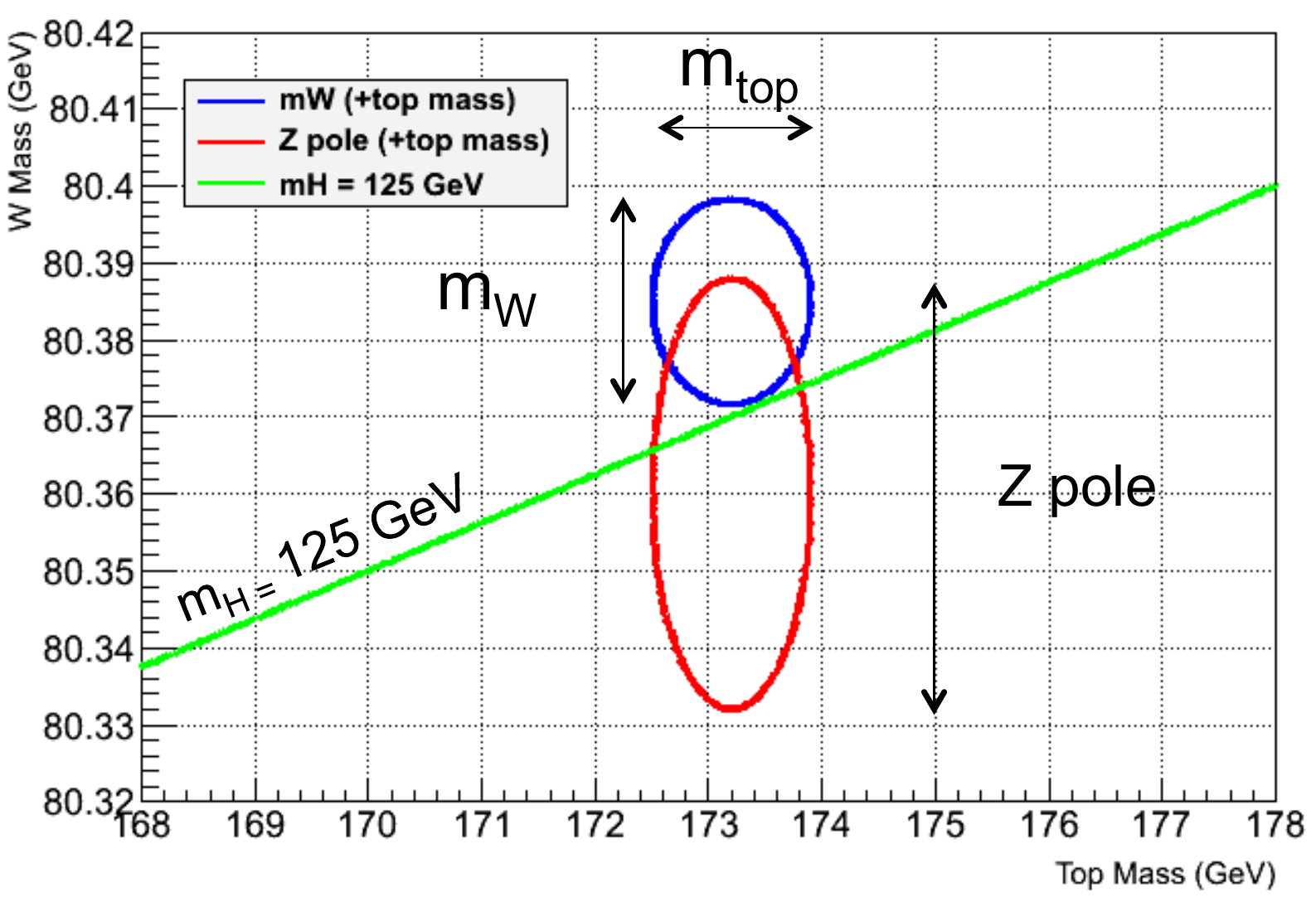}
\includegraphics[width=0.38\textwidth]{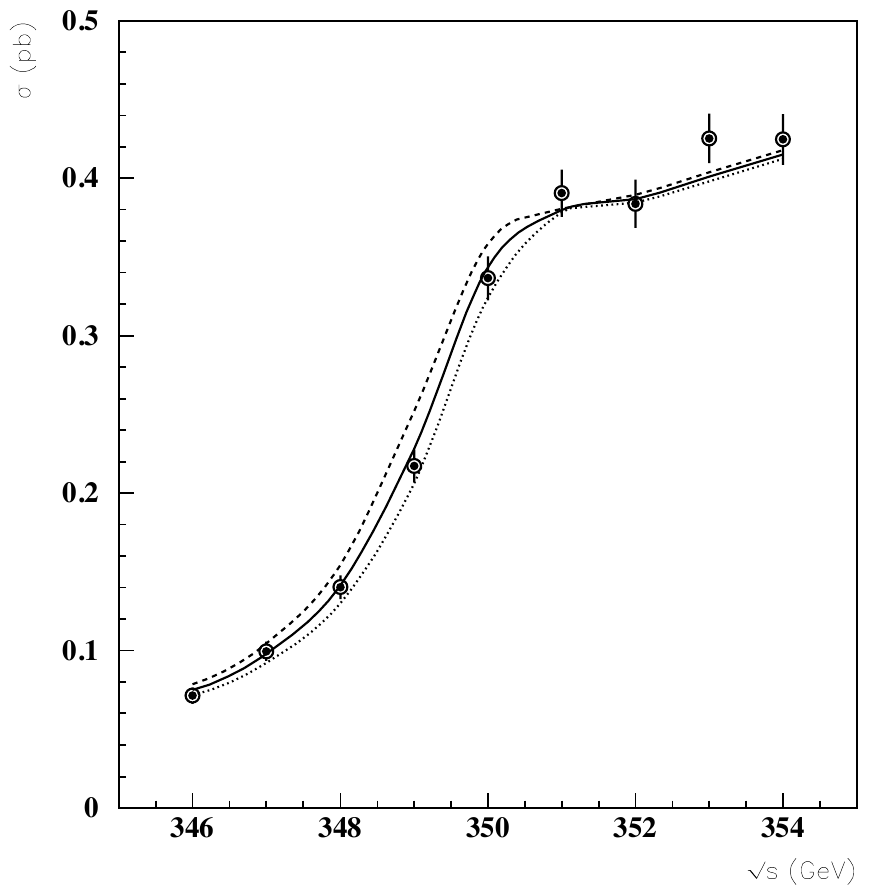}
 \caption{\small (Left) Executive overview of what could be the physics landscape in 2017, just prior to the second long shutdown of LHC. The blue ellipse stands for W-boson and top-quark mass measurements to 10 and 500\,MeV/$c^2$, respectively, the red ellipse shows the W mass prediction from the existing Z pole measurements combined with the top mass knowledge, and the green line is the SM prediction for a Higgs boson mass of 125\,GeV/$c^2$. (Right) A simulated scan of the ${\rm t \bar t}$ threshold at an ${\rm e^+e^-}$ collider with $\sqrt{s} \simeq 350$\,GeV. The dotted curves indicate shifts of the top mass by $\pm 100$\,MeV/$c^2$.}
  \label{fig:mtmw2017}
\end{center}
\end{figure*}

With the complete analysis of the TeVatron data and of the forthcoming LHC data at 13 TeV in 2015-17, the W boson and top quark masses will optimistically be measured with a 10\,MeV/$c^2$ and 500\,MeV/$c^2$ precision, respectively. Together with the expected measurement of the Higgs boson mass with a 100\,MeV/$c^2$ precision, the right graph of Fig.~\ref{fig:mtmw} would become something like the one presented in the left graph of Fig.~\ref{fig:mtmw2017}, where the top mass measurement is combined with the Z pole measurements to predict the W mass in the SM framework. 

Should no additional discoveries be made at the LHC before the second long shutdown in 2018-2019, the scientific grounds to continue the LHC adventure with the high-luminosity programme (HL-LHC) might weaken slightly, and the case of the high-energy version of the LHC (HE-LHC) would be strengthened accordingly. Without any guidance on where or what to look for, the natural next steps would then be to further constrain the Higgs mechanism predictions by reducing drastically the sizes of the blue and red ellipses of Fig.\ref{fig:mtmw2017}, and to measure with high precision all the properties of the Higgs boson. This programme would allow the consistency of the SM to be decisively challenged and, more importantly, it would provide precious information on where new physics hides. 

\begin{figure*}[hbtp]
\begin{center} 
\includegraphics[width=0.48\textwidth]{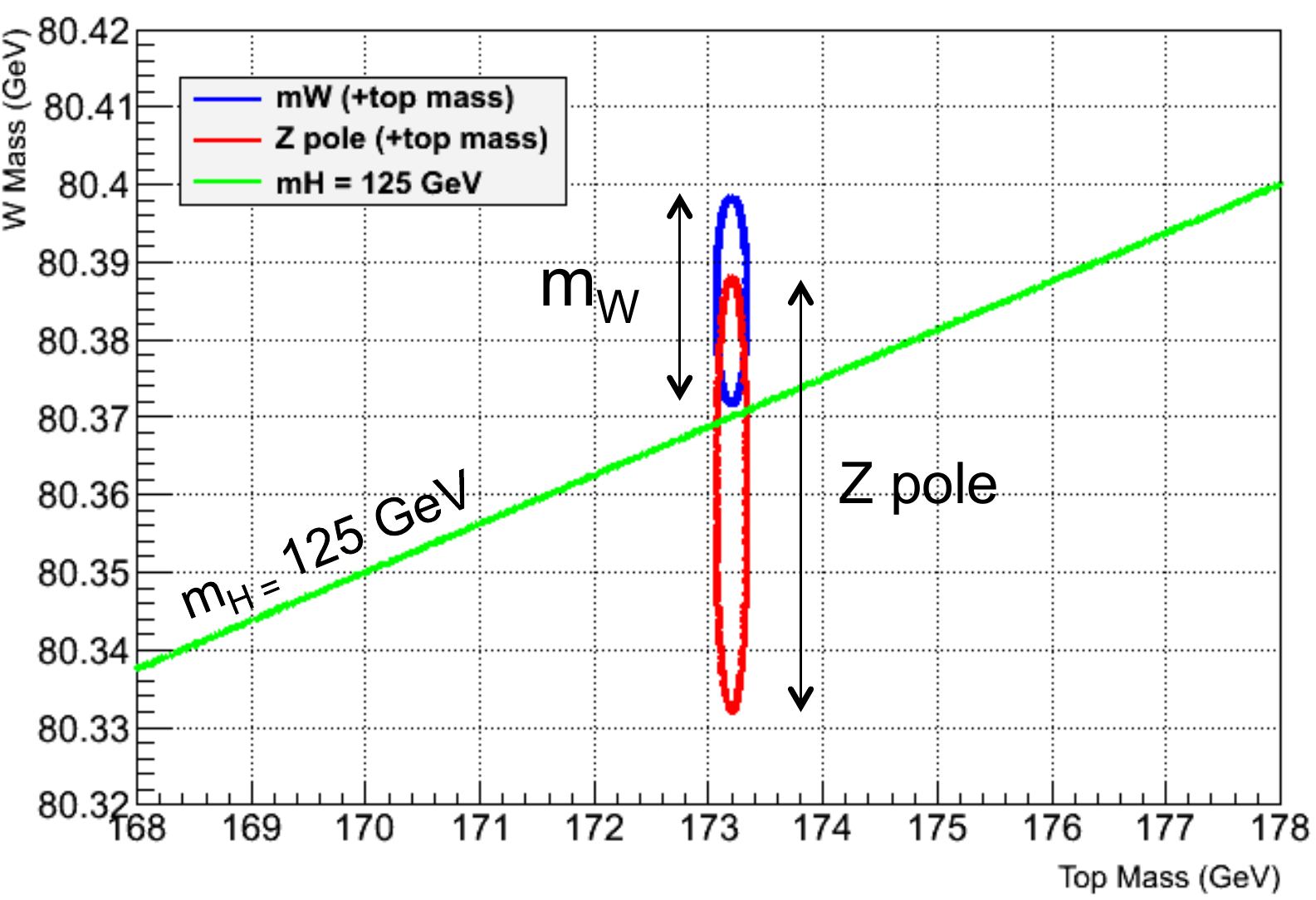}
\includegraphics[width=0.48\textwidth]{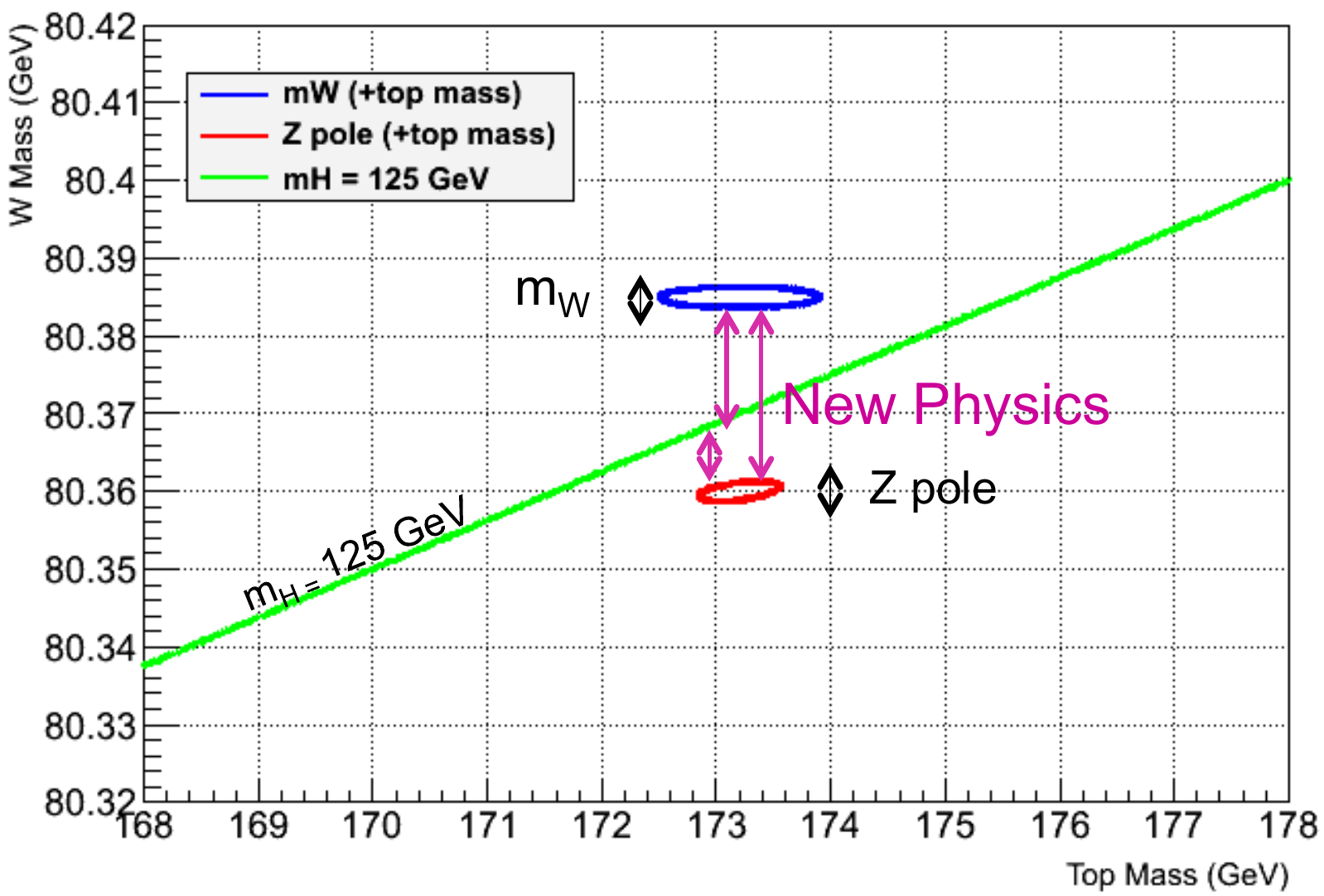}
\caption{\small The effect of improving the precision of the top mass measurement by a factor 10 (left), or the precision of the Z pole and the W mass measurements by factors of 25 and 10, respectively (right), in the $(m_{\rm W}, m_{\rm top})$ plane.}
  \label{fig:mtmwmw}
\end{center}
\end{figure*}

A first possibility along this path would be to measure the top mass with a precision of 50\,MeV/$c^2$, achievable with a scan of the ${\rm t \bar t}$ production threshold, {\it e.g.}, at an ${\rm e^+e^-}$ collider with $\sqrt{s} \simeq 350$\,GeV, as displayed in the right graph of Fig.~\ref{fig:mtmw2017}~\cite{TESLA}.
While measuring the top mass and the top cross section is experimentally genuinely interesting, the SM does not really get immediately challenged by this measurement, as can be seen from the left graph of Fig.~\ref{fig:mtmwmw}: the red and blue ellipses still overlap significantly, and the green line does not provide additional information as to the SM consistency. In addition, the theoretical meaning of the top-quark mass is still being debated, so that this measurement may well be affected by sizeable systematic uncertainties when entering this figure. 

Instead, a improvement by a factor of 10 of the W mass precision (thus reduced to 1\,MeV/$c^2$), and by a factor 25 of the Z pole measurements, which can be achieved by repeating the LEP programme with much larger statistics, would lead to the graph shown in the right of Fig.~\ref{fig:mtmwmw} -- if the central values were not to be modified by the new measurements.  Non-overlapping ellipses would point to new particles coupling to the Z and/or the W, and would provide hints as to the nature and the masses of these new particles. Similarly, ellipses not overlapping with the green line would hint at a non-standard Higgs mechanism. Needless to say, Fig.~\ref{fig:mtmwmw} is given here for illustration only and does not represent the full power of the LEP3 programme at the Z pole: other projections might lead to more hints for new physics.

The LEP3 programme would therefore consists in three different phases (in whichever order) for an overall duration of 5 to 10 years: {\it (i)} precise Z pole measurements at $\sqrt{s} \simeq m_{\rm Z}$, for one year; {\it (ii)} a precise W mass measurement at the WW production threshold, $\sqrt{s} \simeq 2m_{\rm W}$, for one year; and {\it (iii)} a precise Higgs mechanism characterization at $\sqrt{s} = 240$\,GeV, for five years. The characteristics of the LEP3 proposal for each of these three phases are quickly examined in turn in Sections~\ref{sec:HiggsFact} to~\ref{sec:MegaW}.

\subsection{LEP3 as a Higgs factory: \boldmath{$\sqrt{s} = 240$} \,GeV}
\label{sec:HiggsFact}

\subsubsection{Centre-of-Mass Energy}

\begin{figure*}[hbtp]
\begin{center} 
\includegraphics[width=0.48\textwidth]{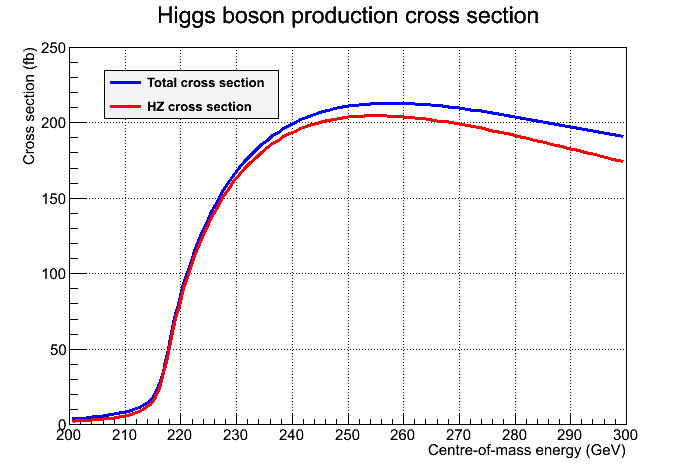}
\includegraphics[width=0.48\textwidth]{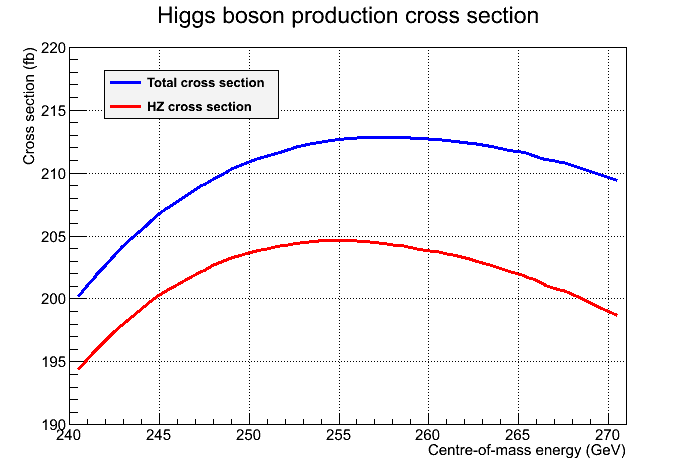}
\caption{\small The Higgs boson production cross section as a function of the centre-of-mass energy. The red curve corresponds to the Higgsstrahlung process only, ${\rm e^+e^-} \to {\rm HZ}$, and the blue curve includes the WW and ZZ fusion processes as well, together with their interference with the Higgsstrahlung process. The right graph is a zoom of the left graph around the maximum of the cross section.}
  \label{fig:HZHA}
\end{center}
\end{figure*}

The centre-of-mass energy of a Higgs factory can be chosen to maximize the Higgs boson production cross section. This cross section, determined with the {\tt HZHA} generator~\cite{HZHA} used by the four LEP collaborations during the LEP2 phase, is displayed in Fig.~\ref{fig:HZHA} as a function of the centre-of-mass energy for $m_{\rm H} = 125$\,GeV/$c^2$.
The maximum cross section of $\sim 212$\,fb, arguably optimal for precision Higgs boson studies, is obtained for $\sqrt{s} \sim 260$\,GeV, and it stays above 200\,fb from 240 to 285\,GeV. Going from 240 to 260\,GeV corresponds to a cross-section increase of 6\%, but would lead to an increase of the power consumption, hence of the operating cost, of 40\%. It was hence decided to make the present studies with $\sqrt{s}= 240$\,GeV.

\subsubsection{Luminosity and Number of Events Expected}

With the most recent parameters of the machine, an instantaneous luminosity of $10^{34}\,{\rm cm}^{-2}{\rm s}^{-1}$ or more can be achieved at this energy in each interaction point, thus providing each detector with an integrated luminosity of $500\,{\rm fb}^{-1}$, {\it i.e.}, a total of 100,000 Higgs bosons to study, over a period of five years. With this integrated luminosity, the numbers of signal and background events expected in each detector are as listed in Table~\ref{tab:events}.
\begin{table*}[htbH]
\begin{center}
\topcaption{\small Numbers of of events expected in each detector for signal (left) and background (right) with $500\,{\rm fb}^{-1}$ at $\sqrt{s}=240$\,GeV. The {\tt HZHA} generator was used for signal, and {\tt PYTHIA} was used for all backgrounds, except for the ${\rm e^+e^-} \to {\rm Z}\nu\bar\nu$ process, for which a private code was developed. For the first two background processes, the ${\rm Z}^*/\gamma^*$ invariant mass was required to exceed 30\,GeV/$c^2$. For the Bhabha process, at least one of the final state electron was required to be in the detector acceptance, {\it i.e.}, with an angle with respect to the beam axis larger than $5^\circ$. For the last process, the $\gamma\gamma$ centre-of-mass energy was required to exceed 2\,GeV.\label{tab:events}}
\begin{tabular}{|l|c|r||l|c|r|c|}
\hline\hline 
Signal & BR (\%) & Events & Background & $\sigma$ (pb) & Events & Rate (Hz) \\
\hline\hline
${\rm H} \to {\rm b\bar b}$ & 57.9 & 57,870 & ${\rm e^+e^-} \to {\rm Z}^*/\gamma^* \to {\rm q\bar q}$ & 50 & 25,000,000 & 0.50 \\
\hline
${\rm H} \to {\rm W^+ W^-}$  & 21.6 & 21,630 & ${\rm e^+e^-} \to {\rm Z}^*/\gamma^* \to \ell^+\ell^-$ & 12.5 & 6,250,000 & 0.12 \\
\hline
${\rm H} \to {\rm gg}$      & 8.19 &  8,200 & ${\rm e^+e^-} \to {\rm W^+W^-}$ & 16 & 8,000,000 & 0.16 \\
\hline
${\rm H} \to \tau^+\tau^-$  & 6.40 &  6,400 & ${\rm e^+e^-} \to {\rm Z Z }$ & 1.3 & 650,000 & 0.01 \\
\hline
${\rm H} \to {\rm c\bar c}$ & 2.83 &  2,820 & ${\rm e^+e^-} \to {\rm We}\nu $ & 1.35 & 700,000 & 0.01 \\
\hline
${\rm H} \to {\rm Z Z}$     & 2.62 &  2,620 & ${\rm e^+e^-} \to {\rm Ze^+e^-}$ & 3.8 & 1,900,000 & 0.04 \\
\hline
${\rm H} \to \gamma\gamma$  & 0.27 &    266 & ${\rm e^+e^-} \to {\rm Z}\nu\bar\nu $ & 0.032 & 16,000 & -- \\
\hline
${\rm H} \to{\rm Z}\gamma$  & 0.16 &    160 & ${\rm e^+e^-} \to {\rm e^+e^-}$ (Bhabha) & 5,000 & $2.5\,10^9$ & 50 \\
\hline
${\rm H} \to  \mu^+\mu^-$   & 0.02 &     22 & $\gamma \gamma \to \ell^+\ell^-, {\rm q \bar q}$ & 15,000 & $7.5\,10^9$ & 150 \\
\hline\hline
\end{tabular}
\end{center}
\end{table*}

\subsubsection{Pile-up and Beamstrahlung}
The rates for the Bhabha process and the $\gamma\gamma$ interactions do not exceed 200\,Hz. With $4\times 4$ colliding bunches, the beam crossing rate at each interaction point is 45\,kHz, {\it i.e.}, more than 200 times larger than the rate of interactions potentially producing signal in the detector. The probability of pileing-up either a Bhabha event or a $\gamma\gamma$ interaction with a signal event is therefore smaller than 0.5\%. This favourable situation is unique to a collider ring. In a linear collider, the luminosity is obtained with extremely small vertical beam sizes (5\,nm for ILC and 1\,nm for CLIC, to be compared to 320\,nm for LEP3) because the collision rate is substantially smaller (typically 5\,Hz for ILC and 50\,Hz for CLIC). These small collision rates would lead to 40 and 4 pile-up events per pulse, respectively. The quite-relaxed bunch time structure of ILC (2820 bunches separated by 337 ns in each pulse) allows this pile-up rate to be reduced to 1.5\%, but the large drift time of a TPC would integrate the hits of two or three $\gamma\gamma$ interactions on top of each interesting event. The bunch time structure of CLIC (312 bunches separated by 0.5\,ns in each pulse) might be more delicate to handle in this respect~\cite{pileup}.

The comfortable vertical beam size of LEP3 also leads to more favourable beamstrahlung conditions. With the current machine parameters, the resulting beam energy spread is expected to be smaller than 0.1\%, {\it i.e.}, much smaller than the spread of 2\% expected from initial state radiation, and much smaller than the beamstrahlung-induced energy spread expected at linear colliders (3.2\% for ILC, and somewhat larger for CLIC). It is to be noted that the photons from beamstrahlung may also increase significantly the rate of $\gamma\gamma$ interactions at a linear collider, hence the pileup rate.

\subsubsection{Detectors}

Another advantage of a collider ring is the possibility of recording data with more than one detector. Besides the larger integrated luminosity useable for physics, which scales with the number of operating detectors, this possibility has also favourable sociological and geo-political consequences. While CMS and ATLAS would be the natural first choices if the cost of the project is to be kept as low as possible, nothing would prevent from adding other two detectors based on the technologies developed for linear colliders by different experimental collaborations. These configurations with either two or four detectors are chosen to draw the final conclusions of this note.

\subsubsection{Integrated Luminosity Measurement}

Focusing quadrupoles need to be installed very close to the interaction points, typically at a distance of about 4\,m. Very small prototypes, with an aperture of 5\,cm and a Focusing gradient of 17\,T/m, have already been produced at CERN in the context of the CLIC R\&D. These quadrupoles would typically cover an angular region of less than 5 degrees around the beam axis, where precision luminometers are usually placed. The Bhabha cross section, however, amounts to 1\,nb if the two electrons are to be detected in the region from 5 to 20 degrees. In this region, usually covered by regular electromagnetic calorimeters, the contamination from the $s$ channel and the $s$-$t$ channel interference is less than 0.1\%. A total of 500 million Bhabha events is expected with an integrated luminosity of 500\,${\rm fb}^{-1}$, making the statistical uncertainty on the integrated luminosity not an issue for the purpose of this note. 

\subsubsection{Beam Energy and W Mass Measurement}

The beam energy can be determined from the precise knowledge of the Z mass with two processes: {\it (i)} the ${\rm e^+e^-} \to \gamma{\rm Z}$ process; and {\it (ii)} the ${\rm e^+e^-} \to {\rm ZZ}$ process. With 500\,${\rm fb}^{-1}$, each detector would collect one million $\gamma$Z events (with ${\rm Z} \to {\rm e^+e^-}, \mu^+\mu^-$) and 400,000 ZZ events (with none of the two Zs decaying into $\nu\bar\nu$). With techniques similar to those developed to measure the W mass at LEP2 from WW production, a statistical uncertainty of 5\,MeV on the average beam energy can be obtained with ZZ production for each detector. The measurement of the beam energy with $\gamma$Z production was studied in detail Ref.~\cite{returns}. With one million such events, a statistical uncertainty of 3\,MeV is achievable. A combination of these two measurements performed with four detectors can lead to an ultimate precision of 1\,MeV. 

It is interesting to note that, with a sample of eight million W pairs per experiment, the statistical uncertainty on the W mass measurement would also be of the order of 1\,MeV/$c^2$ (for each experiment), of the same order as the systematic uncertainty arising from the aforementioned beam energy knowledge. This uncertainty would lead to the blue ellipse of Fig.~\ref{fig:mtmwmw}.

\subsection{LEP3 as a Tera-Z factory: \boldmath{$\sqrt{s} \simeq m_{\bf Z}$}}
\label{sec:TeraZ}

\subsubsection{Luminosity and Rates}

With the RF power needed to keep four bunches in the LEP3 ring at $\sqrt{s}=240$\,GeV, it is possible to have 50 times more current in the machine at $\sqrt{s} \sim m_{\rm Z}$, distributed in 200 bunches. As a consequence, an instantaneous luminosity as large as $5\times 10^{35}\,{\rm cm}^{-2}{\rm s}^{-1}$ can be delivered at the Z peak, to be compared to a few $10^{33}\,{\rm cm}^{-2}{\rm s}^{-1}$ at a linear collider. It is important to realize that, since the charge in each bunch is the same as in each of the four bunches at the highest centre-of-mass energy, beamstrahlung and pileup effects are not more of an issue at the Z peak than they are at the Higgs factory, in spite of the much larger instantaneous luminosity.

With 5\,${\rm ab}^{-1}$ per year and per experiment, a sample of $\sim 10^{12}$ Zs can be accumulated in one year, thus opening an entirely new field of precision measurements. For the record, LEP1 accumulated a sample of about 20 million Zs. The LEP1 programme could be repeated every 10 minutes, a quite useful feature for detector calibration. All statistical errors can therefore be potentially reduced by more than two orders of magnitude with LEP3, thus producing a red ellipse four times smaller than that of Fig.~\ref{fig:mtmwmw}. As at higher energy, the integrated luminosity can be measured with Bhabha scattering in the region from 5 to 20 degrees from the beam axis. In this region, the Bhabha cross section amounts to 10\,nb at $\sqrt{s} = m_{\rm Z}$, with less than 0.5\% contribution from the Z, leading to $2\times 10^{11}$ ${\rm e^+e^-}$ events in a year. To further increase the statistics, a precision luminometer could also be installed just in front of the focussing quadrupoles and cover the angular region from 1 to 5 degrees from the beam axis. In any case, specific studies need to be undertaken to establish whether the detector accuracy and the theoretical predictions can match the statistical power of this sample. 

Another important point to realize is that, with this luminosity, the rate of interesting events (Z decays + Bhabha events + $\gamma\gamma$ interactions, to only cite the dominant contributors) will reach 25\,kHz or thereabout. This rate is 25 times larger than the 1\,kHz of events currently written onto disk by the CMS experiment operated at LHC. Luckily, both the size and the processing time of the LEP3 events are over 20 times smaller than those of the current CMS events, arising from quite busy proton-proton collisions and superimposed with an average of 30 minimum bias events. The existing high-level trigger farm of CMS could therefore probably cope with an output rate of 25\,kHz in ${\rm e^+e^-}$ collisions at the Z peak.

\subsubsection{Beam Energy Measurement}

A technique unique to ${\rm e^+e^-}$ rings, called resonant spin depolarization~\cite{depolarization}, will be used to measure the beam energy with high precision. This technique was developed and successfully used during the LEP1 programme, and allowed the beam energy to be known with a precision of 2\,MeV. The intrinsic precision of the method, 0.1 MeV or better, was not attained at LEP1 because the polarization could not be maintained in collisions. The few measurements performed in specific runs with one beam had therefore to be extrapolated to ``predict'' the beam energy during collisions, with different conditions. This extrapolation was the dominant contributor to the 2\,MeV uncertainty.

At LEP3, instead, it will be possible to inject a few non-colliding bunches in addition to the 200 colliding bunches, and apply resonant spin depolarization on those. The beam energy will therefore be measured continuously, in the exact same conditions as for the colliding bunches. A beam energy uncertainty of 0.1\,MeV or better, {\it i.e.}, with a $10^{-6}$ relative precision, is therefore at hand, crucial to measure the Z width with the required accuracy.

\subsubsection{Polarization} 

At LEP1, it was possible to establish a 60\% polarization with a single 45 GeV electron beam. With some care in the design of the LEP3 ring, it might be possible to maintain this high level of polarization in collisions at the Z peak. The Z peak cross section is proportional to $1-P_+P_-+A_{\rm LR}(P_++P_-)$, where $P_+$ and $P_-$ are the average longitudinal polarizations of the colliding positron and electron bunches, respectively, and $A_{\rm LR} = 8(1-\sin^2\theta_{\rm W})$ is the left-right asymmetry. A scheme with alternate polarized and non-polarized bunches proposed in Ref.~\cite{polarization} would allow a simultaneous measurement of the beam polarization and the left-right asymmetry $A_{\rm LR}$ with unprecedented precision.  

\subsection{LEP3 as a Mega-W factory: \boldmath{$\sqrt{s} \simeq  2m_{\bf W}$}}
\label{sec:MegaW}

As part of the programme, LEP3 could be operated at $\sqrt{s}=2m_{\rm W}$ to measure the WW cross section at threshold, $\sigma_{\rm WW}$, and to derive the W mass from this measurement. At this centre-of-mass energy, LEP3 will be able handle about 10 times more current than at $\sqrt{s}=240$\,GeV, distributed in 40 bunches. The instantaneous luminosity would then amount to $10^{35}\,{\rm cm}^{-2}{\rm s}^{-1}$, corresponding to an integrated luminosity of 1\,${\rm ab}^{-1}$ in a year. The WW cross section of $\sim 4$\,pb would then lead to the production of four million W pairs in each experiment. 

With only $10\,{\rm pb}^{-1}$ per experiment, a combined precision of 220\,MeV/$c^2$ was achieved on the W mass at LEP1 from the WW cross section measurement. The $10^5$ larger data sample foreseen at LEP3 would reduce the statistical precision to 0.7\,MeV/$c^2$. Radiative returns to the Z, ${\rm e^+e^-} \to \gamma$Z with ${\rm Z} \to {\rm e^+e^-}, \mu^+\mu^-$, will provide a measurement a the beam energy to a similar precision. This beam-energy measurement will possibly be backed-up by resonant spin depolarization, provided that beam polarization can be established at these energies (which was never achieved at LEP2, although theoretically feasible). In that case, a combined precision on the W mass of the order of 0.3\,MeV/$c^2$ would be achieved with four detectors.

\subsubsection{A side remark: TLEP}
As mentioned above, new discoveries at the LHC after the first long shutdown could put LEP3 in competition with the HL-LHC part of the LHC programme, scheduled to start after the second long shutdown. To avoid such conflicts, a longer term programme at CERN can be envisioned in a new, 80\,km-long, tunnel, which could host first a 350\,GeV ${\rm e^+e^-}$ collider ring, called TLEP, later followed by a very-high-energy pp collider ($\sqrt{s}=100$\,TeV) in the same tunnel. The LEP3 physics programme, described above, can of course be done at TLEP with no modification, albeit with five times larger instantaneous luminosities at $\sqrt{s} = 240$\,GeV, hence with more than twice better accuracy on all Higgs boson branching fractions. The physics programme would then be completed by a five-year run at the ${\rm t\bar t}$ threshold for a precise determinatation of the top-quark properties and decays. The cost of TLEP would remain significantly smaller than that of a linear collider. 

\section{Precision measurements with CMS at the Higgs Factory}
\label{sec:CMS}

\subsection{The CMS Apparatus}

The CMS detector~\cite{cmsdetector} is performing extremely well in proton-proton and heavy-ion collisions. Very quickly, it appeared almost ideally suited to reconstruct and identify all individual stable particles (photons, muons, electrons, charged and neutral hadrons) arising from a collision, with the so-called particle-flow event reconstruction~\cite{PFT-09-001,PFT-10-001,PFT-10-002,PFT-10-003}. With its large silicon tracker immersed in a uniform axial magnetic field of 3.8\,T provided by a superconducting solenoidal coil, charged-particle tracks can be reconstructed with large efficiency and adequately small fake rate down to a momentum transverse to the beam ({\PT}) of 200\,MeV/$c$, all the way to $20^\circ$ from the beam axis. 

Photons are reconstructed with an excellent energy resolution ($\sigma_E/E = 3.5\%/\sqrt{E} \oplus 0.3\%$) by an essentially hermetic electromagnetic  calorimeter (ECAL) surrounding the tracker and made of over 75,000 2\,cm$\times$2cm PbWO$_4$ crystals, down to $5^\circ$ from the beam axis. Together with the large magnetic field, this fine ECAL granularity is a key element, as it generally allows photons to be separated from charged-particle energy deposits even in jets with an energy of several hundred GeV. The granularity is enhanced by an order of magnitude in the end-caps by the presence of a lead and silicon-strip pre-shower (ES) placed in front of the ECAL crystals. 

Charged and neutral hadrons deposit their energy in the hadron calorimeter (HCAL) made of brass and scintillators, installed inside the coil and surrounding the ECAL, with a similar angular coverage. The granularity of the HCAL is 25 times coarser than that of the ECAL, which still allows charged and neutral hadrons to be spatially separated in jets with an energy below 100\,GeV. The hadron energy resolution in the combined ECAL--HCAL system, of the order of 10\% at 100\,GeV, also allows neutral hadrons to be detected as an energy excess on top of the energy deposited by the charged hadrons pointing to the same calorimeter cells. The charged hadrons are reconstructed with the superior angular and energy resolutions of the tracker, combined with those of the calorimeter system. In proton-proton collisions, particles between 1 and $5^\circ$ from the beam axis are more coarsely measured with an additional forward calorimeter (HF), placed 11\,m from the interaction point. 

This forward calorimeter, however, will be masked by the focusing quadrupoles needed to get the luminosity in ${\rm e^+e^-}$ collisions. The information coming from this detector is therefore ignored in the present work. As already mentioned, a precision luminometer might be placed in the same angular region, in front of the focusing quadrupoles. The study of this possibility is beyond the scope of this note.

Electrons are reconstructed by a combination of a track and of several energy deposits in the ECAL,  from the electron itself and from possible Bremsstrahlung photons radiated by the electron in the tracker material on its way to the ECAL. Muons are reconstructed and identified, in isolation as well  as in jets, with very large efficiency and purity from a combination of the tracker and muon chamber information. The resolution on the Z boson mass in ${\rm e^+e^-}$ ($\mu^+\mu^-$) is 1.5 (1.2)\,GeV/$c^2$ in the central part of the detector. 

The presence of neutrinos and other weakly-interacting particles (for example from a Higgs boson decaying to a neutralino pair) can be detected by the missing momentum $p_{\rm miss}$, defined as the negative vector sum of the momenta of all reconstructed particles, and the missing energy $E_{\rm miss}$, defined as the difference between the centre-of-mass energy $\sqrt{s}$ and the sum of the energies of all reconstructed particles. The missing energy resolution can be parameterized as $\sigma_{E_{\rm miss}} = 45\% E_{\rm miss} + 0.55$\,GeV.  

Finally, the reconstructed particles are clustered in jets with the anti-$k_{\rm T}$ algorithm~\cite{antikt}, operated with a distance parameter of 0.5. The jet energy resolution can be parameterized as $\sigma_E/E = 50\%/\sqrt{E} \oplus 4\%$ and the jet angular resolution is about 30\,mrad for energies below 100\,GeV. The hadron-plus-strips (HPS) algorithm~\cite{hps} is used to identify hadronically-decaying taus, by considering jets with either one charged pion and up to two neutral pions, or three charged pions. The Combined Secondary Vertex (CSV) b-tagging algorithm~\cite{btaga} is used to identify jets that are likely to arise from the hadronization of b quarks. This algorithm combines the information about track impact parameters and secondary vertices within jets in a likelihood discriminant to provide separation of b jets from jets originating from light quarks and gluons, and charm quarks. Several working points for the CSV output discriminant (with values between 0. and 1.) are used in the analysis, with different efficiencies and misidentification rates for b jets. For a ${\rm CSV} > 0.90$ requirement the efficiencies to tag b quarks, c quarks, and light quarks, are approximately 50\%, 6\%, and 0.15\%, respectively~\cite{btagb}. The corresponding efficiencies for ${\rm CSV} > 0.50$ are approximately 72\%, 23\%, and 3\%.

The signal and background processes of Table~\ref{tab:events} were simulated with the GEANT-based simulation of the CMS detector~\cite{PTDR1}. The detector hardware geometry and the reconstruction software used for the present analysis were kept unchanged with respect to those tuned for pp collisions with 30 pileup collisions in each bunch crossing. This choice was purposely made to give here a conservative estimate of the CMS performance in ${\rm e^+e^-}$ collisions at LEP3, although substantial optimizations can be envisioned. Similarly, very simple selection algorithms were developed to reject the various backgrounds and to analyse the Higgs boson properties, leaving room for significant future improvements. Whenever justified, the list of the possible developments and the subsequent potential gains is given in the following sections. Finally, not all the Higgs boson decays are yet addressed hereafter. The examples given are therefore to be taken as an illustration of the possible achievements, and by no means represent a full account of the ultimate LEP3 and CMS performance towards the Higgs boson characterization. More final states will be studied by the end of the year, and the present note will be updated accordingly. In the following sections and in all related figures, it is assumed that LEP3 will be running for five years as a Higgs factory at the nominal instantaneous luminosity.

\subsection{Measurement of the \boldmath{${\rm e^+e^-} \to {\rm HZ}$} cross section, \boldmath{$\sigma_{\rm HZ}$}}
\label{sec:hz}

The ${\rm e^+e^-} \to {\rm HZ}$ cross section, $\sigma_{\rm HZ}$, can be measured in a model-independent manner with the ${\rm Z}\to{\rm e^+e^-}, \mu^+\mu^-$ decays, without any selection on the recoiling system. This measurement is a direct probe of the $g_{\rm HZZ}$ coupling. As Higgs bosons are predominantly produced through this process at $\sqrt{s} = 240$\,GeV, the precision of all other couplings is limited by the accuracy of this cross section measurement.

\begin{figure*}[hbtp]
\begin{center} 
\includegraphics[width=0.48\textwidth]{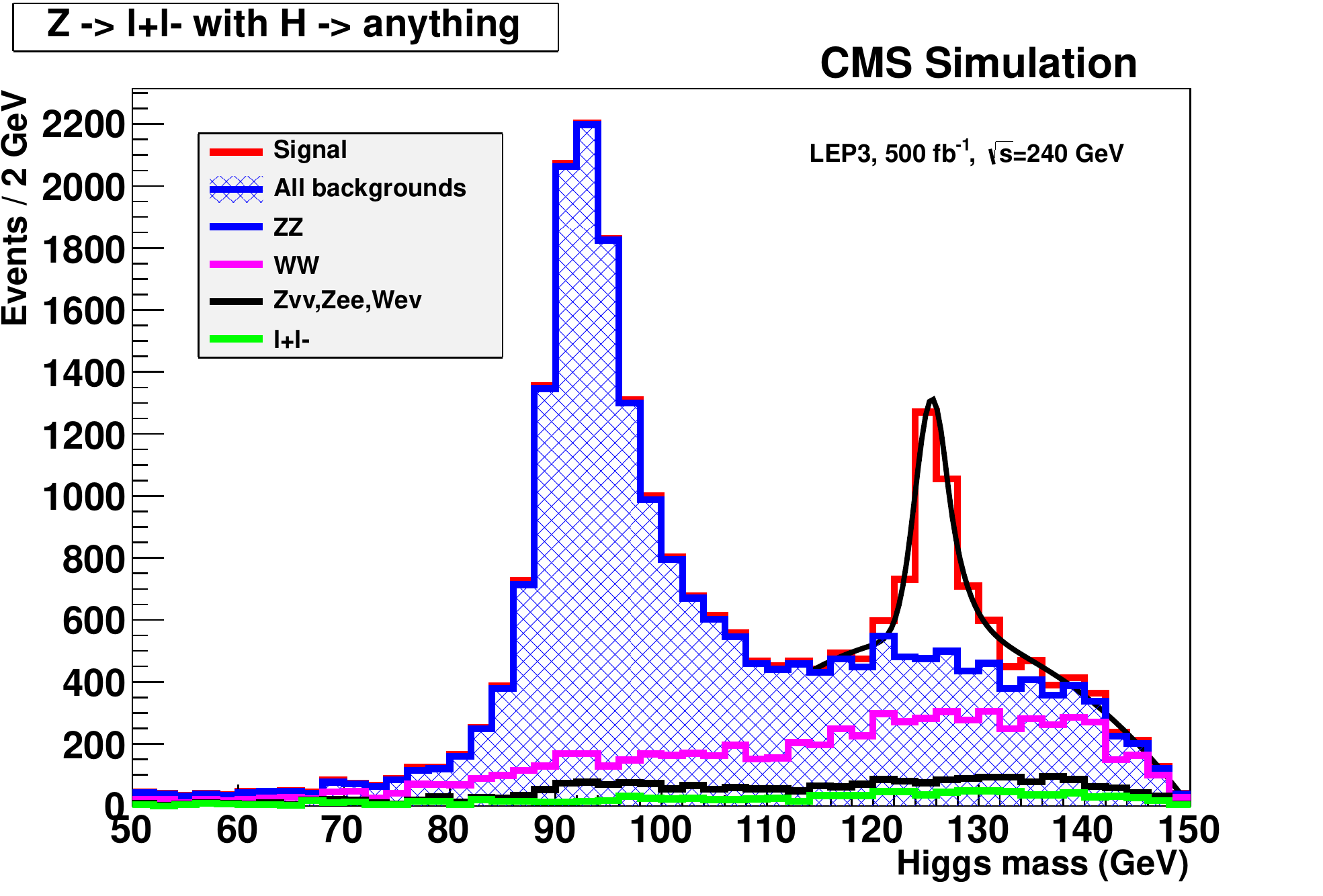}
\includegraphics[width=0.48\textwidth]{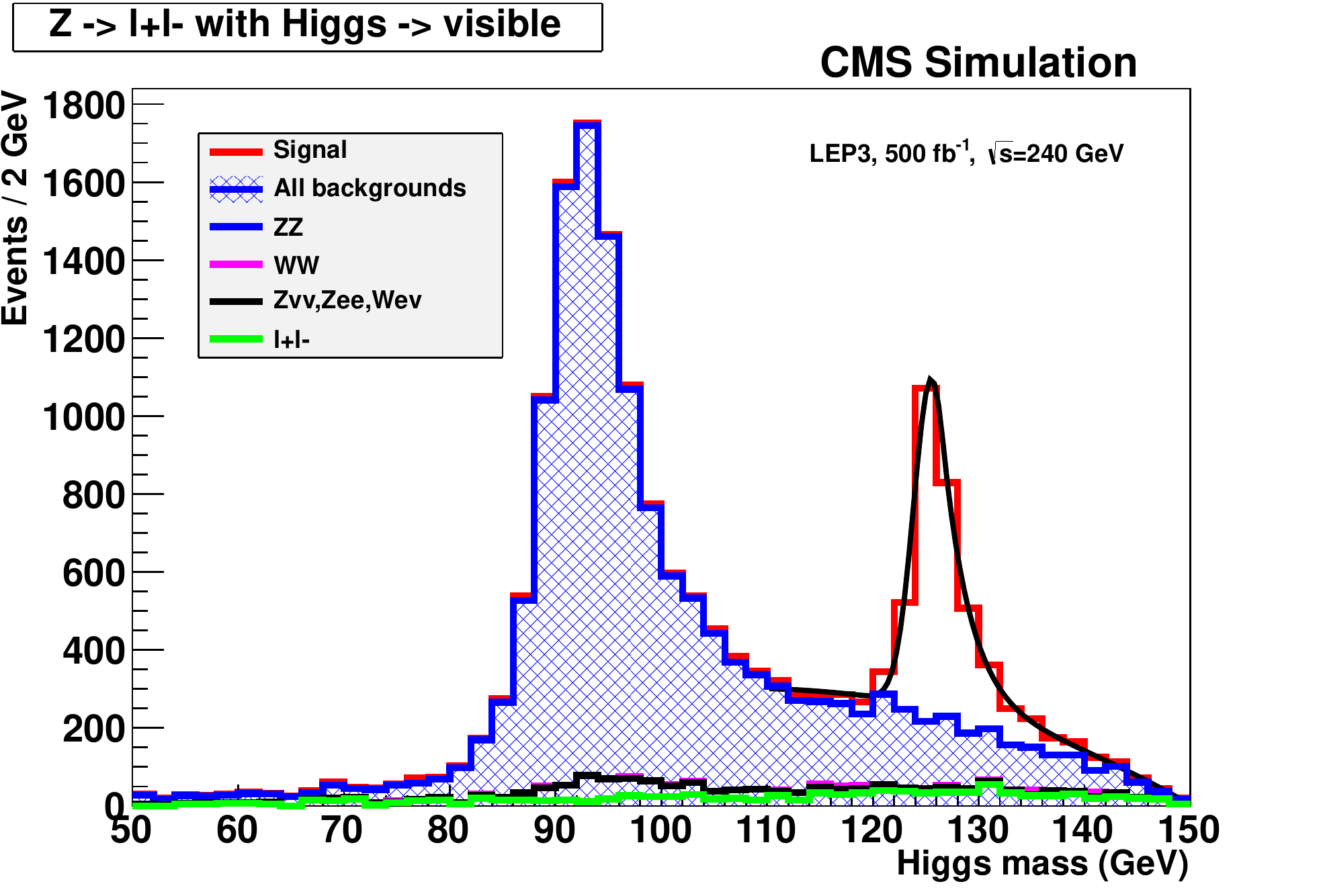}
\caption{\small Distribution of the mass recoiling against the lepton pair (labelled ``Higgs mass'') in HZ candidate events with ${\rm Z}\to{\rm e^+e^-}, \mu^+\mu^-$ for the HZ signal (hollow histogram) and all backgrounds (shaded histogram). Left: no cuts is applied on the recoiling system; Right: the recoiling system is required to be visible, {\it i.e.}, to consist of two reconstructed ``jets''. (These jets may contain only one particle just above the detection thresholds.)}
\label{fig:llX}
\end{center}
\end{figure*}

Events with an oppositely-charged same-flavour lepton pair are selected. The energy of the candidate Brems\-strahlung photons detected in the vicinity of the leptons is added to the closest-lepton energy, and the  invariant mass of lepton pair is required to be compatible with the Z mass, within $\pm 5$\,GeV/$c^2$. To reject $\ell^+\ell^- (\gamma)$ events with a radiative return to the Z mass and a photon lost along the beam axis, the transverse momentum of the lepton pair is required to exceed 10\,GeV/$c$, its longitudinal momentum must be smaller than 50\,GeV/$s$, and the acoplanarity, defined as the angle between the lepton pair plane and the beam axis, must be larger than $10^\circ$. Radiative returns to the Z mass with a photon in the detector acceptance are dealt with by vetoing events with one additional jet carrying more than 80\% of electromagnetic energy. Finally, the ZZ background is reduced by requiring the acollinearity, defined as the angle between the two leptons, to be larger than $100^\circ$. The distribution of the mass recoiling against the lepton pair is shown in Fig.~\ref{fig:llX} (left), with no additional cuts on the recoiling system, for the signal and all backgrounds.
 
The signal is fit to a Crystal Ball function and the backgrounds to a third order polynomial. A precision of 3.1\% on $\sigma_{\rm HZ}$ can be achieved with this fit. A multivariate analysis would probably reduce the background by a factor of two, and improve the precision to $\sim 2.7\%$. Another possibility is to measure the visible branching fraction of the Higgs boson, by imposing the recoiling system to form two distinct ``jets'' -- even if consisting of only one particle -- in the detector. The WW background is almost entirely rejected with this additional criterion, as is shown in the right graph of Fig.~\ref{fig:llX}, and a precision of 2.6\% is obtained on $\sigma_{\rm HZ} \times {\rm BR(H}\to{\rm visible})$. By constraining the invisible decay branching fraction of the Higgs boson, as is exemplified in the next section, the precision obtained on  $\sigma_{\rm HZ}$ becomes 2.7\%.

\subsection{Measurement of the invisible decay branching fraction of the Higgs boson}

\begin{figure*}[hbtp]
\begin{center} 
\includegraphics[width=0.48\textwidth]{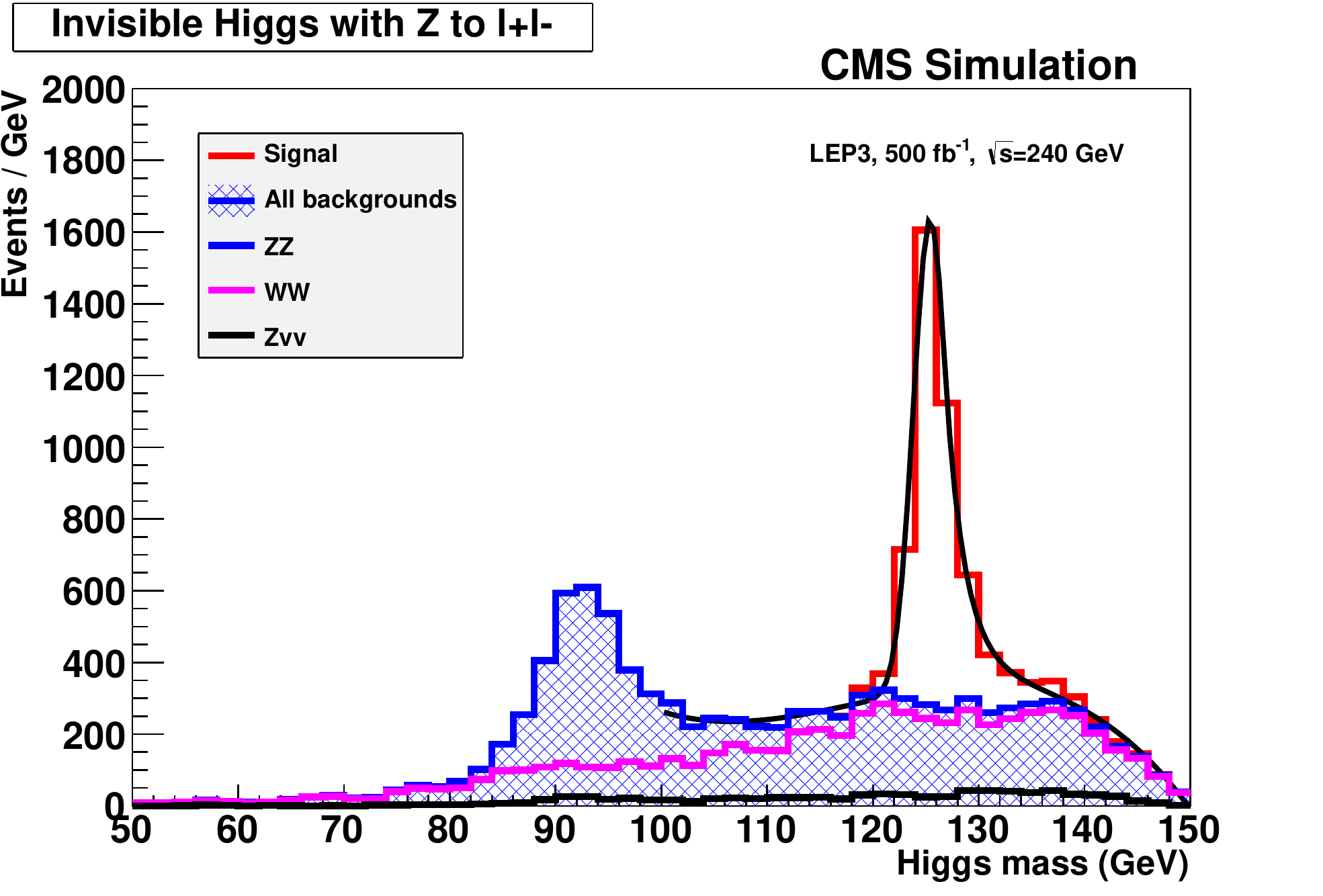}
\includegraphics[width=0.48\textwidth]{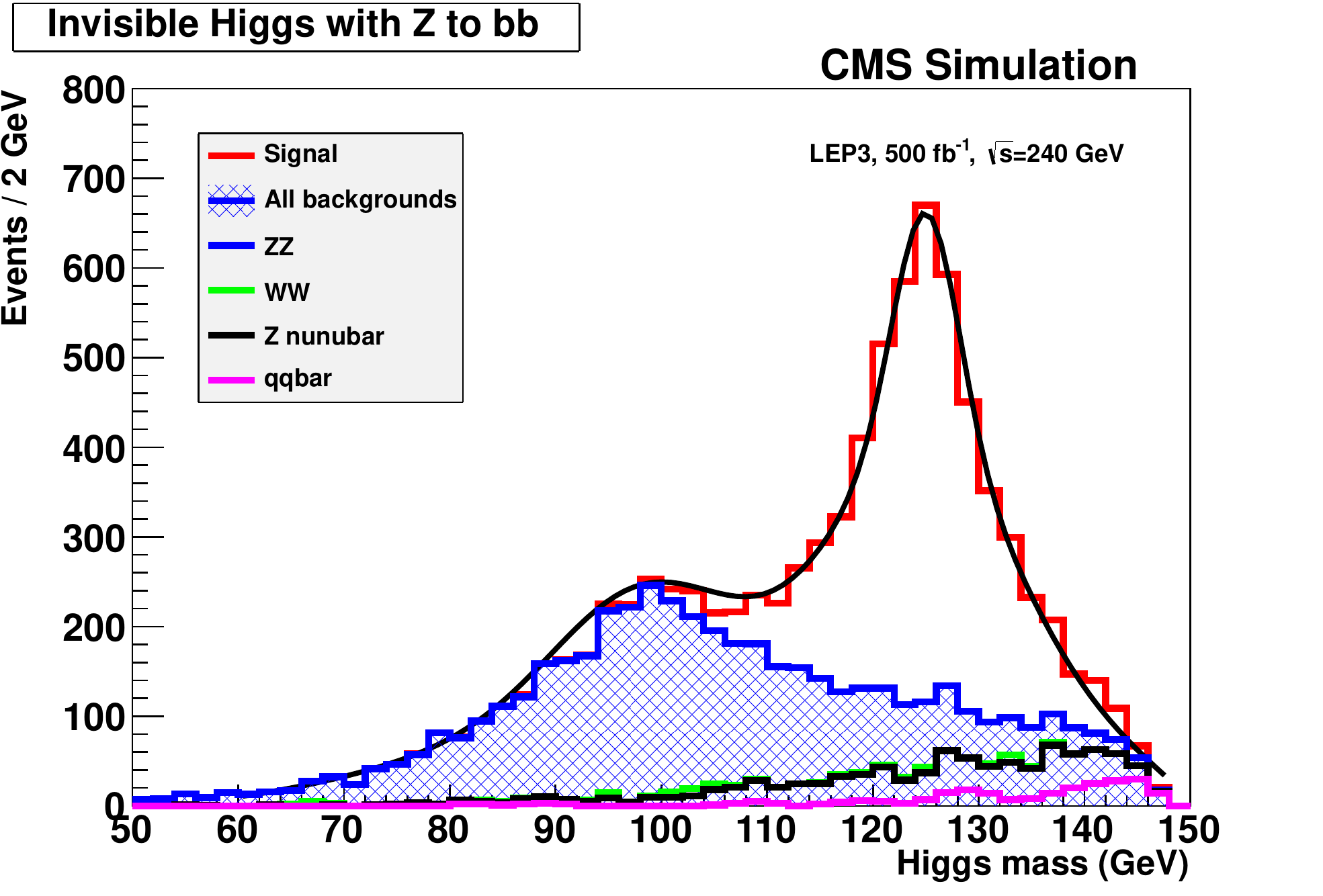}
\caption{\small Distribution of the mass recoiling against the lepton pair (labelled ``Higgs mass'') in events with only an ${\rm e^+e^-}$ or a $\mu^+\mu^-$ pair (left) and with events with only a pair of b-tagged jets (right), for the HZ signal (hollow histogram) and all backgrounds (shaded histogram). Here, the Higgs boson is assumed to decay invisibly with a 100\% branching ratio.}
\label{fig:invis}
\end{center}
\end{figure*}

The invisible branching fraction of the Higgs boson (which would include the decay into a pair of neutralinos, or any other dark matter candidate) can be measured in a very similar manner by selecting events with only two ``jets'', each containing just an electron or a muon and possible Bremsstrahlung photons. Cuts very similar to those listed in Section~\ref{sec:hz} are applied to the lepton pair, and the lepton momenta are fit to the Z mass constraint. The distribution of the mass recoiling to the lepton pair is displayed in the left graph of Fig.~\ref{fig:invis}, for a invisible Higgs decay branching fraction of 100\%. In this configuration, the cross section $\sigma_{\rm HZ}$ can be measured with a precision of 2.2\%.

The combined precision reduces to 1.4\% with a second selecion of events with two b-tagged jets coming from the decay of the Z, with proceeds as follows. Events with less than two jets are rejected. Each event with more than two jets is forced to form two jets, with a recombination the original jets of particles, starting with the pair of smallest mass and iterating. The sum of the two-jet ${\rm CSV}$ b-tagging variables must exceed 0.95 and the dijet invariant mass is required to be between 70 and 100\,GeV/$c^2$. Finally, the same cuts on the dijet acollinearity, acoplanarity, transverse momentum and longitundinal momentum as in the lepton pair case are applied. The distribution of the mass recoiling to the b-jet pair is displayed in the right graph of Fig.~\ref{fig:invis}, for a invisible Higgs decay branching fraction of 100\%. A $5\sigma$ observation is still expected if the invisible decay branching ratio amounts to 4\%. In case of non-observation of this decay mode, an invisible branching fraction of 1.5\% can be excluded at the 95\% confidence level. (In these last two estimates, the cross-contamination of the visible Higgs boson decays via ${\rm e^+e^-} \to {\rm HZ} \to {\rm b\bar b}\nu\bar\nu$ is taken into account.)

\subsection{Measurement of \boldmath{$\sigma_{\rm HZ} \times {\rm BR(H}\to{\rm b\bar b})$}}
\label{sec:bb}

The ${\rm H \to b \bar b}$ decay can be measured in three different channels: {\it (i)} the leptonic channel, addressing the ${\rm ZH} \to \ell^+\ell^-{\rm b \bar b}$ decay; {\it (ii)} the four-jet channel, addressing the ${\rm ZH} \to {\rm q \bar q b \bar b}$ decay; and {\it (iii)} the missing energy channel, addressing the $\nu\bar \nu ({\rm H \to b \bar b})$ final state.

\subsubsection{The leptonic channel}

\begin{figure*}[hbtp]
\begin{center} 
\includegraphics[width=0.60\textwidth]{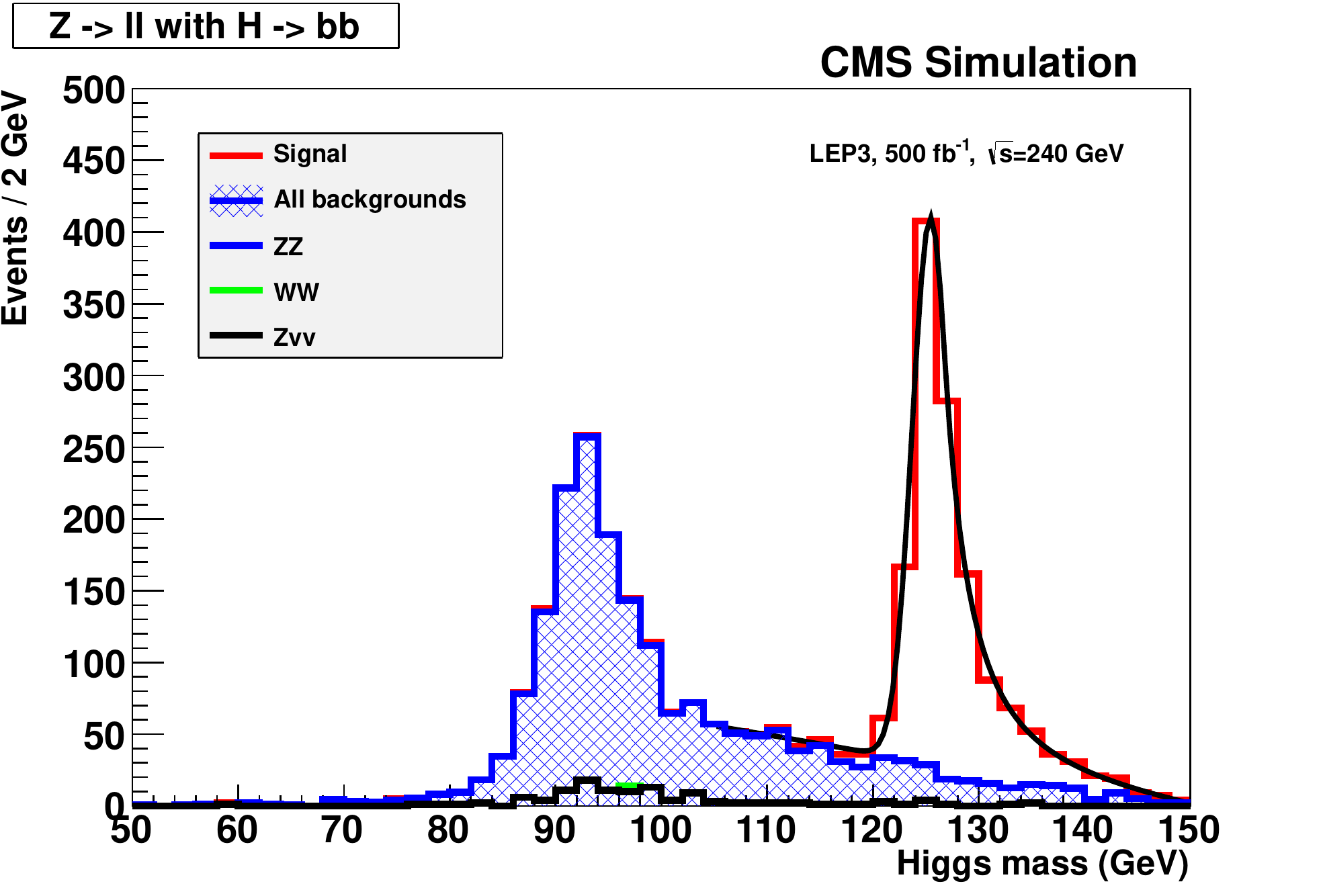}
\caption{\small Distribution of the mass recoiling against the lepton pair (labelled ``Higgs mass'') in the leptonic channel, for the HZ signal (hollow histogram) and all backgrounds (shaded histogram).}
\label{fig:llbb}
\end{center}
\end{figure*}

The exact same selection as for the $\sigma_{\rm HZ}$ measurement is applied, with the additional requirement that the sum of the largest two {\rm CSV} b-tagging variables of the recoiling system be larger than 0.90. The distribution of the mass recoiling to the lepton pair is shown in Fig.~\ref{fig:llbb}, and is fit to the sum of Crystal Ball function for the signal and a third order polynomial for the backgrounds. A precision of 3.1\% is achieved on the $\sigma_{\rm HZ} \times {\rm BR(H}\to{\rm b\bar b})$ measurement.

In the range from 120 to 130\,GeV/$c^2$, the 950 signal events selected are 99\% pure in ${\rm H \to b\bar b}$ decays. When the b tagging criterion is reversed, the resulting signal event sample of about 1200 events still consists of ${\rm H \to b\bar b}$ decays for 50\%, but is enriched in other decays, ${\rm H \to c\bar c}$, gg, ${\rm W^+W^-}$ and ZZ. These events can be further classified in several subsamples according to the number of jets and the b-quark content, each with different compositions in ${\rm c\bar c}$, gg, ${\rm W^+W^-}$ and ZZ decays. For example, the sample made of events with at least four jets and a CSV value smaller than 0.1 contains only 4\% of the ${\rm H \to c\bar c}$ events, but 30\% of the ${\rm H \to W^+W^-}$ or ZZ events and 25\% of the ${\rm H \to gg}$ events. In contrast, the sample made of events with less than four jets and a CSV value between 0.1 and 0.9 contains 50\% of the ${\rm H \to c\bar c}$ events, but only 5\% of the ${\rm H \to W^+W^-}$ or ZZ events and 10\% of the ${\rm H \to gg}$ events.

By solving the corresponding set of linear equations, a statistical accuracy of 18\% is obtained for $\sigma_{\rm HZ} \times {\rm BR(H}\to{\rm c\bar c\ or\ gg})$. A more precise determination of this quantity would require the commissioning and the use of specific c-tagging and gluon-tagging algorithms, yet not available at CMS.

\subsubsection{The four-jet channel}

To select the ${\rm ZH} \to {\rm q \bar q b \bar b}$, events with less than four jets are rejected, and events with more than four jets are forced to form four jets, by iteratively recombining the jet pair with the smallest invariant mass. To reject Z decays into leptons (e, $\mu$ and $\tau$ pairs), all jets must have at least five reconstructed particles, of which at least one charged hadron. To reject Z decays into a neutrino pair as well as other events with missing energy, the visible mass is required to exceed 180\,GeV/$c^2$. The four jet energies are then rescaled to ensure the total energy/momentum conservation, keeping the jet velocities fixed. All rescaled energies must be positive and the $\chi^2$ defined as the sum of the four $(E_{\rm rescaled}-E_{\rm measured})^2/\sigma_E^2$ is required to be smaller than 15. The compatibility with fully hadronic ZZ and WW events is measured with $\Delta_{\rm WW,ZZ}$ defined as
$$\Delta_{\rm WW,ZZ} = \min_{\{ij,kl\}} \left[ (m_{ij} - m_{\rm W,Z})^2 + (m_{kl} - m_{\rm W,Z})^2 \right]^{1/2} ,$$
where $i,j,k,l$ are the four jet indices. Both $\Delta$s are  required to be larger than 10\,GeV/$c^2$. 

Candidate Higgs bosons are defined as jet pairs with a mass $m_{12}$ larger than 100\,GeV/$c^2$, provided that the other jet pair invariant mass $m_{34}$ be between 80 and 110\,GeV/$c^2$. If more than one jet pair satisfy this criterion, only the one with the largest CSV sum is kept, and the CSV sum is required to exceed 0.95. The reconstructed Higgs boson mass is defined as $m_{\rm H} = m_{12}+m_{34}-m_{\rm Z}$, which gives a 2.8\,GeV/$c^2$ core resolution. The distribution of $m_{\rm H}$ is shown in the left graph of Fig.~\ref{fig:qqbb}, and is fit to the sum of two Gaussian for the signal and a third order polynomial for the background. A precision of 1.5\% is reached on the $\sigma_{\rm HZ} \times {\rm BR(H}\to{\rm b\bar b})$ measurement.

\subsubsection{The missing energy channel}

\begin{figure*}[hbtp]
\begin{center} 
\includegraphics[width=0.48\textwidth]{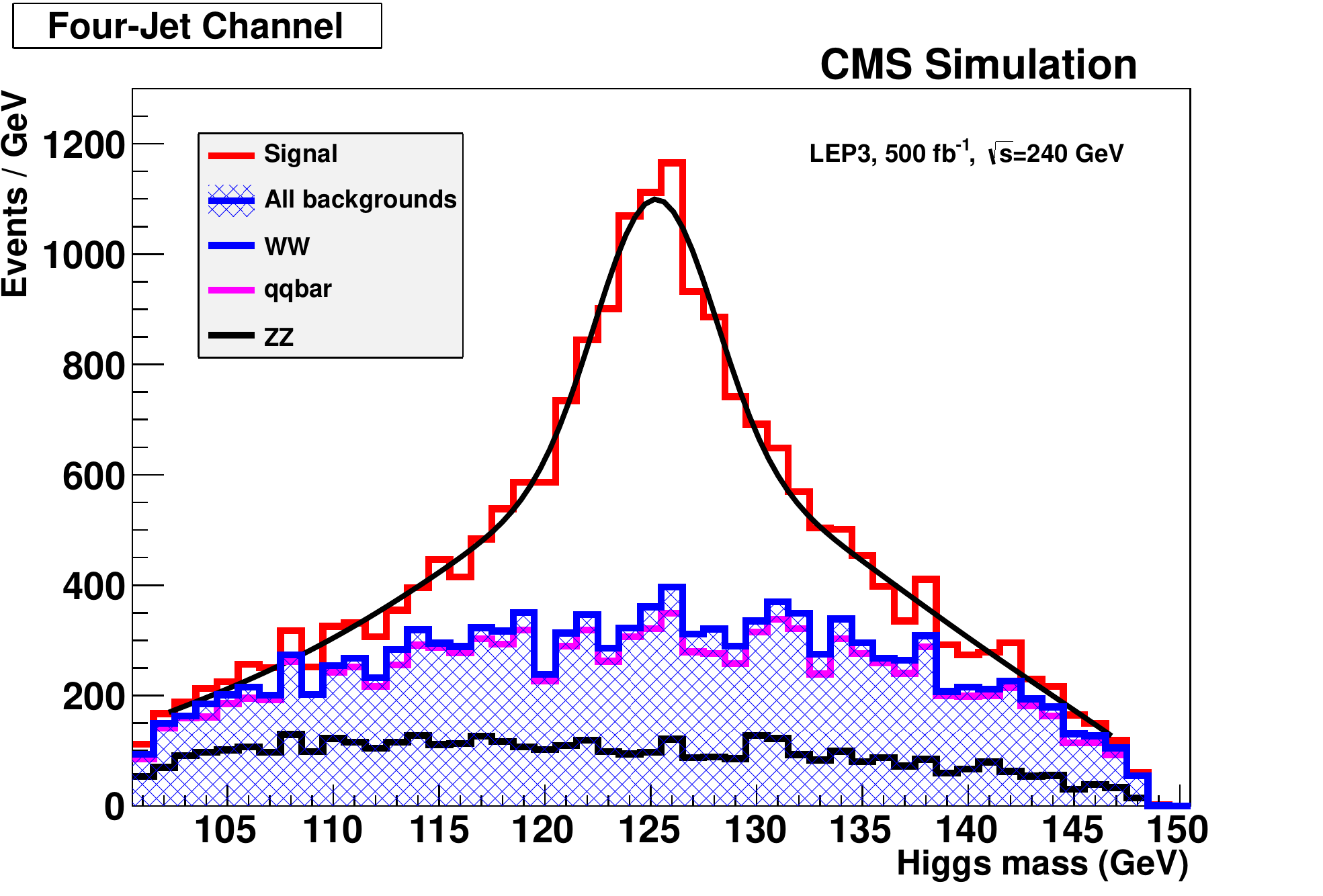}
\includegraphics[width=0.48\textwidth]{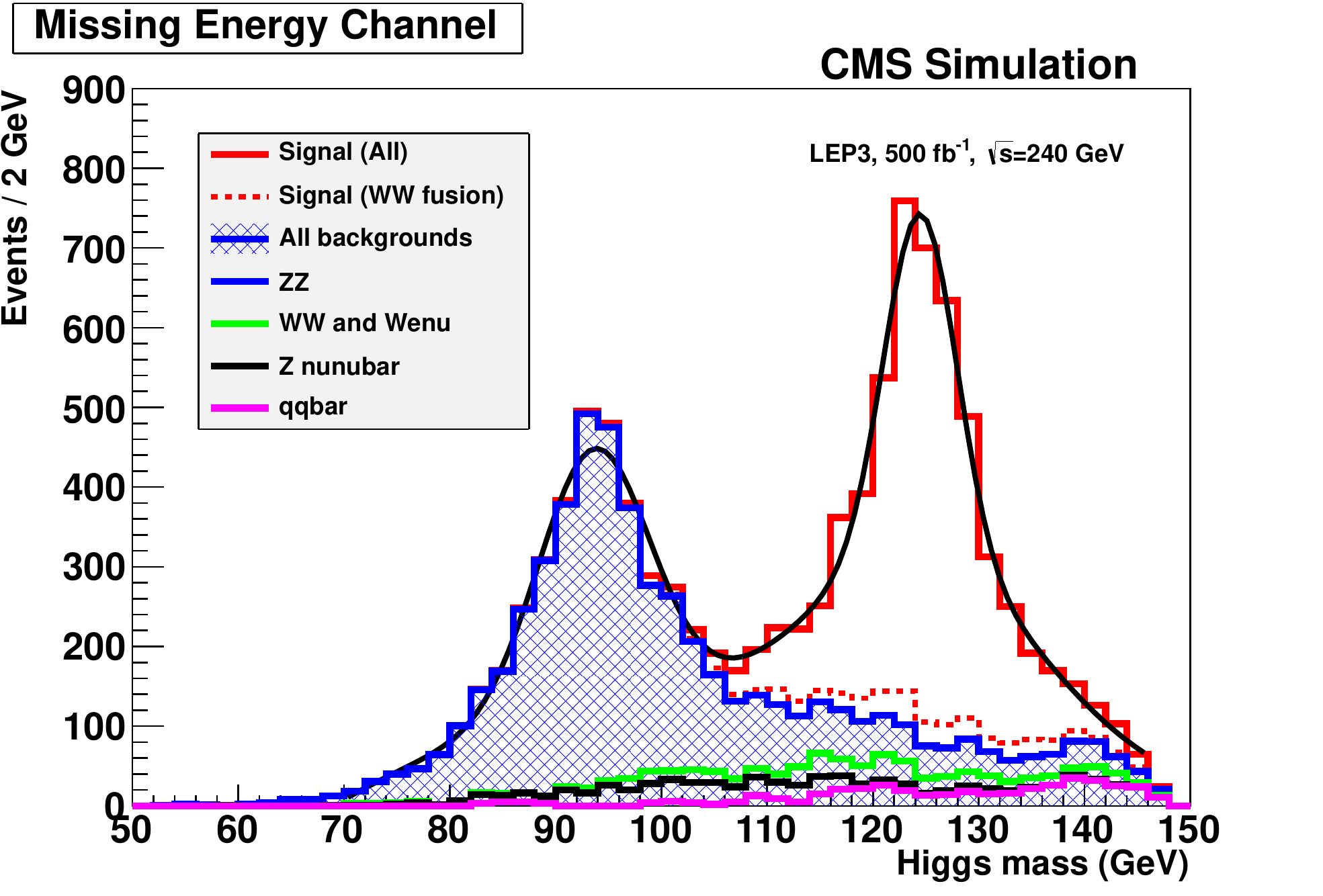}
\caption{\small Distribution of the reconstructed Higgs boson mass in the four-jet channel (left) and in the missing energy channel (right), for the Higgs boson signal (hollow histogram) and all backgrounds (shaded histogram). In the missing energy channel, the contribution from the ${\rm WW \to H}$ fusion process is indicated with a dashed curve.}
\label{fig:qqbb}
\end{center}
\end{figure*}

The selection of the $\nu\bar \nu ({\rm H \to b \bar b})$ events, which arise both from the Higgsstrahlung process with a cross section of 38\,fb and from the WW fusion with a cross section of 7\,fb, proceeds in a way very similar to the selection of the invisible Higgs boson decays accompanied with ${\rm Z \to b\bar b}$. The exact same cuts on the di-jet acollinearity, acoplanarity, transverse momentum, longitudinal momentum and b tagging are applied, and the missing mass $m_{\rm miss}$ is requested to be between 65 and 125\,GeV/$c^2$. To improve the resolution on the Higgs boson mass, the two jet energies are rescaled by a common factor $\alpha$ chosen to satisfy the constraint $m_{\rm miss}^{\rm rescaled} = m_{\rm Z}$. The distribution of the reconstructed Higgs boson mass, $\alpha m_{\rm b \bar b}$, is displayed in the right graph of Fig.~\ref{fig:qqbb}. The signal is fit to the sum of two Gaussian, with a core resolution of 3.4\,GeV/$c^2$. A precision of 1.8\% is achieved on the $\sigma_{\rm HZ} \times {\rm BR(H}\to{\rm b\bar b})$ measurement.

\subsubsection{Combination and improvements}

The combination of the three above measurements returns the $\sigma_{\rm HZ} \times {\rm BR(H}\to{\rm b\bar b})$ value with a precision of 1.0\%. 

This precision is limited in particular by the tagging efficiency of the two b jets from the ${\rm H \to b \bar b}$ decay, which amounts to 30\% for the high-purity cut chosen in the three analyses. The potential b-tagging performance improvements are numerous. For example, the combination with soft-lepton b-tagging, available in CMS, but not used here, increases the single-jet b-tag efficiency by 10\%. The upgraded version of the pixel detector, with one additional layer and twice less material (to be installed just after the first long shutdown) is expected to increase this efficiency by up to 20\% per jet. All in all, a 50\% increase of the ${\rm H \to b \bar b}$ selection efficiency, with the same light-quark background rejection, is at hand, which could improve the expected precision on $\sigma_{\rm HZ} \times {\rm BR(H}\to{\rm b\bar b})$ to 0.8\%.

In the four-jet channel, more improvements are expected to come from the use of a 5C kinematic fit (with the Z mass constraint in addition to the energy-momentum conservation constraints), and even of 6C kinematic fits to identify the ZZ and WW background. A very significant reduction of the Higgs mass RMS resolution with the 5C fit, and of the WW and ZZ background with the 6C fits, are expected here, towards an even better combined precision.

Finally, the missing energy and the four-jet channels can be used as well as the leptonic channel for a measurement of $\sigma_{\rm HZ} \times {\rm BR(H}\to{\rm c\bar c})$  and of $\sigma_{\rm HZ} \times {\rm BR(H}\to{\rm gg})$, conservatively with a progressive loosening of the b-tagging criterion, or better, with a combination a b-tagging improved algorithm and dedicated c-tagging and gluon-tagging algorithms, not yet available at CMS. 

\subsection{Measurement of \boldmath{$\sigma_{\rm HZ} \times {\rm BR(H}\to \tau^+\tau^-)$}}
\label{sec:htt}

\subsubsection{The ${\rm ZH} \to \ell^+\ell^-\tau^+\tau^-$ channel}

Events with a pair of same-flavour, opposite-charge, leptons are selected as described in Section~\ref{sec:hz}. For the events with more than one such lepton pair, the pair with mass closest to the Z mass is chosen, and the difference with the nominal Z mass is required to be smaller than 6\,GeV/$c^2$. The rest of the event must consist of only two jets (thereafter called ``tau''), both containing at most four charged particles. The energies of the two leptons and the two taus are rescaled to ensure the total energy-momentum conservation as done for the four-jet channel. To ensure four-body compatibility, hence reduce the contamination from $({\rm Z \to \ell^+\ell^-})({\rm H \to WW})$ events with the two Ws decaying leptonically, the four rescaled energies are required to be in excess of 5 GeV, and the rescaled $\tau\tau$ mass should agree with the mass recoiling against the lepton pair within 10\,GeV/$c^2$. As the mass recoiling against the lepton pair has better resolution, it is chosen as the reconstructed Higgs boson mass. Finally, to reject ${\rm ZZ \to 4e}$, 2e2$\mu$ and 4$\mu$ events, the missing mass in the event is required to larger than 10\,GeV/$c^2$ if the tau pair contains an electron or a muon. 

The distribution of the reconstructed Higgs boson mass is shown in  Fig.~\ref{fig:ttqq} (left). It is fit to a Crystal Ball function for the signal and a third order polynomial for the background. A precision of 8.4\% is achieved on the $\sigma_{\rm HZ} \times {\rm BR(H}\to \tau^+\tau^-)$ measurement.

\subsubsection{The ${\rm ZH} \to {\rm q\bar q} \tau^+\tau^-$ channel}

Events from the ${\rm ZH} \to {\rm q\bar q} \tau^+\tau^-$  channel must have at least four jets, of which two compatible with arising from either hadronic or leptonic tau decays. To reject ZZ events with one Z decaying to ${\rm e^+e^-}$ or  $\mu^+\mu^-$, the leading lepton (if any) in the tau jets is required to have an energy smaller than 50\,GeV. The pairs of tau candidates with three charged particles each are not considered. If several tau pairs are found in a given event, all pairs are tested by forcing the rest of the event to form two hadronic jets and by rescaling the tau and jet energies to ensure energy-momentum conservation. The combination with the di-jet invariant mass $m_{\rm jj}$ closest to the Z mass is retained, provided that $m_{\rm jj}$ is between 70 and 110\,GeV/$c^2$. The two hadronic jets are required to contain at least three particles of which one charged, and to have a mass larger than 1.5\,GeV/$c^2$. The angle between the two hadronic jets must be larger than $108^\circ$. 

\begin{figure*}[hbtp]
\begin{center} 
\includegraphics[width=0.48\textwidth]{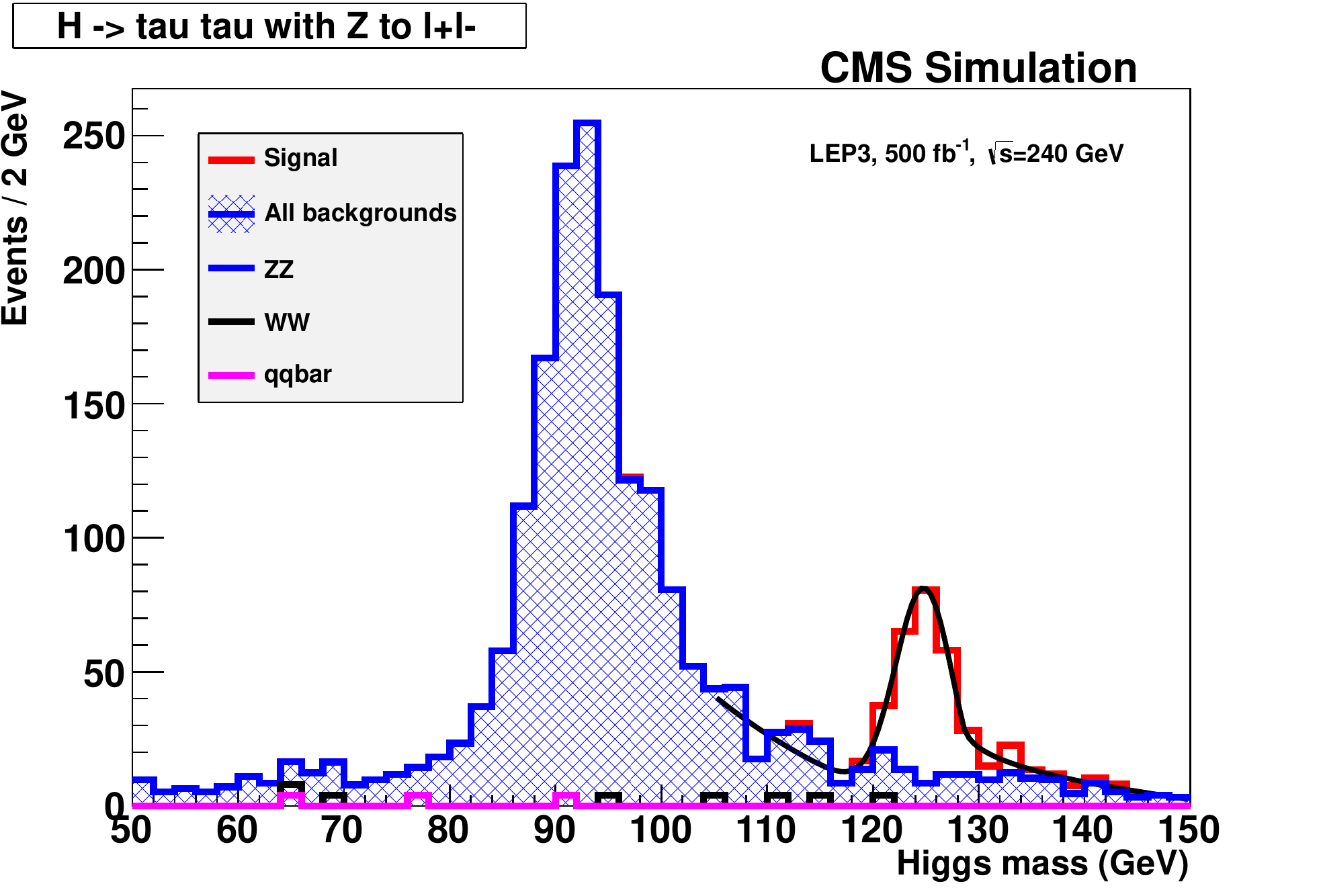}
\includegraphics[width=0.48\textwidth]{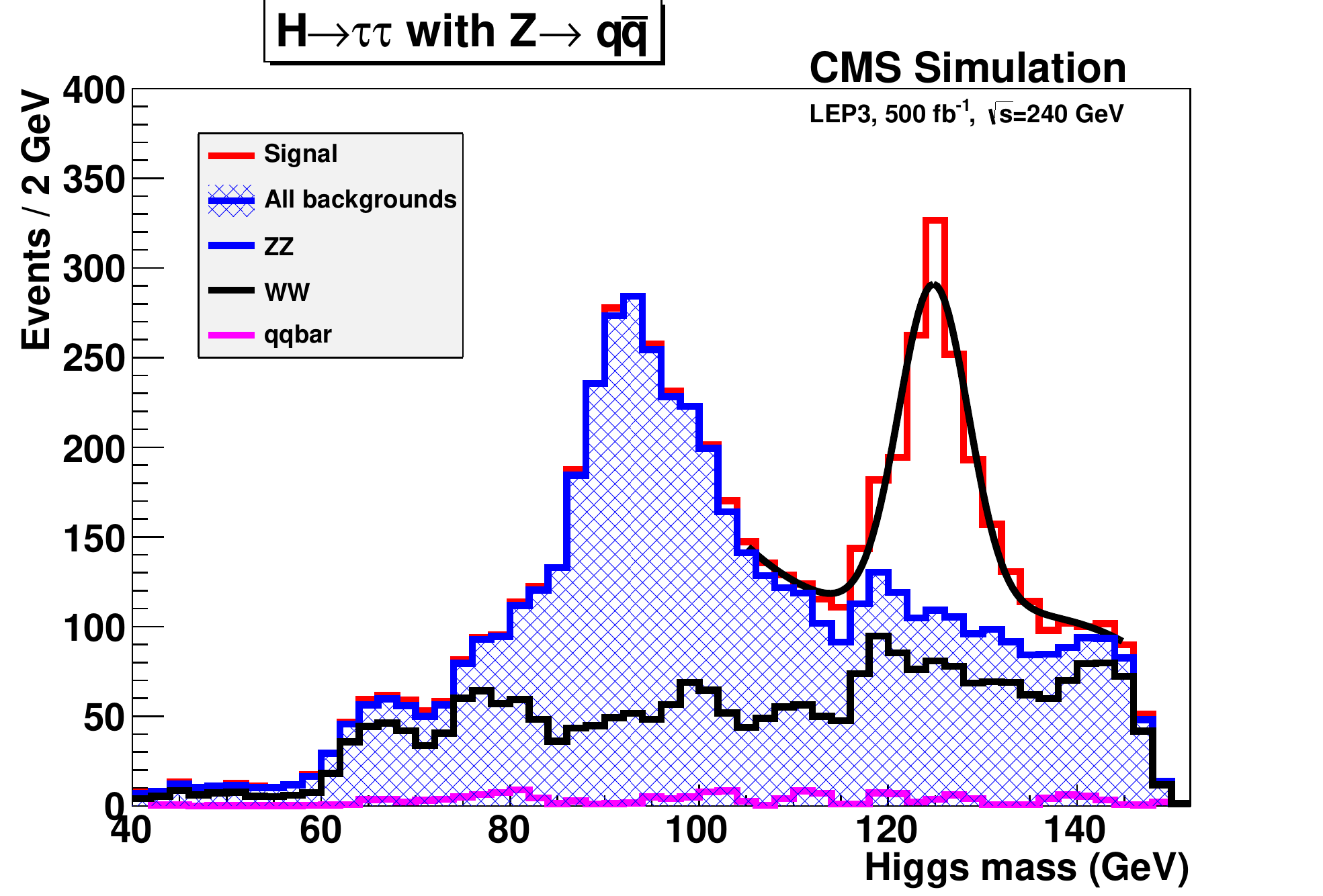}
\caption{\small Distribution of the reconstructed Higgs boson mass (labelled ``Higgs mass'') in the  ${\rm ZH} \to \ell^+\ell^-\tau^+\tau^-$ channel (left) and in the ${\rm ZH} \to {\rm q\bar q}\tau^+\tau^-$ channel (right), for the HZ signal (hollow histogram) and all backgrounds (shaded histogram).}
\label{fig:ttqq}
\end{center}
\end{figure*}

The distribution of the reconstructed Higgs boson mass, defined as $m_{\rm H} = m_{\tau\tau}+m_{\rm jj}-m_{\rm Z}$, is shown in Fig.~\ref{fig:ttqq} (right). The distribution is fit to the sum of a Gaussian for the signal and a third order polynomial for the background. A precision of 5.0\% is achieved on the $\sigma_{\rm HZ} \times {\rm BR(H}\to \tau^+\tau^-)$ measurement. When combined with the ${\rm ZH} \to \ell^+\ell^-\tau^+\tau^-$ channel, the precision becomes 4.3\%.

\subsection{Measurement of \boldmath{$\sigma_{\rm HZ} \times {\rm BR(H}\to {\rm W^+W^-})$}}

\subsubsection{Channels with leptonic Z decays}
\label{sec:2l4q}

Events with an oppositely-charged same-flavour lepton pair are selected as indicated in Section~\ref{sec:hz}. The dilepton invariant mass must be compatible with the nominal Z mass within $\pm 15$\,GeV/$c^2$, and the angle between the two leptons must exceed 110 degrees. The system recoiling to the lepton pair is required to contain at least ten charged hadrons. The fully hadronic decays of the W pair are selected by requiring the recoiling system to consist of at least four jets within the tracker acceptance. The semi-leptonic decays of the W pair are selected by requiring at least one additional electron or muon with momentum smaller than 50\,GeV/$c$, one additional jet in the tracker acceptance, and a total missing transverse momentum in excess of 15\,GeV/$c$.

\begin{figure*}[hbtp]
\begin{center} 
\includegraphics[width=0.48\textwidth]{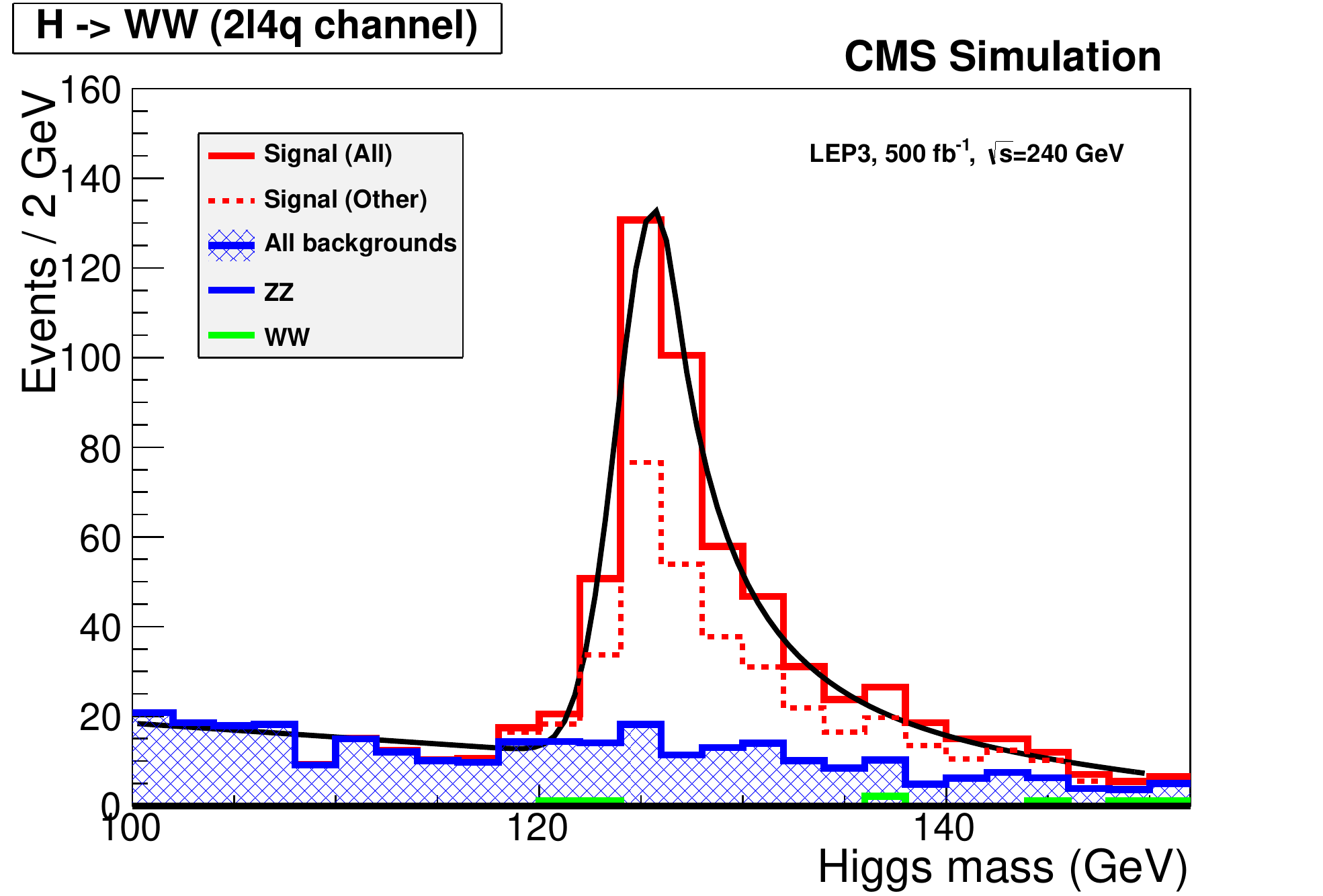}
\includegraphics[width=0.48\textwidth]{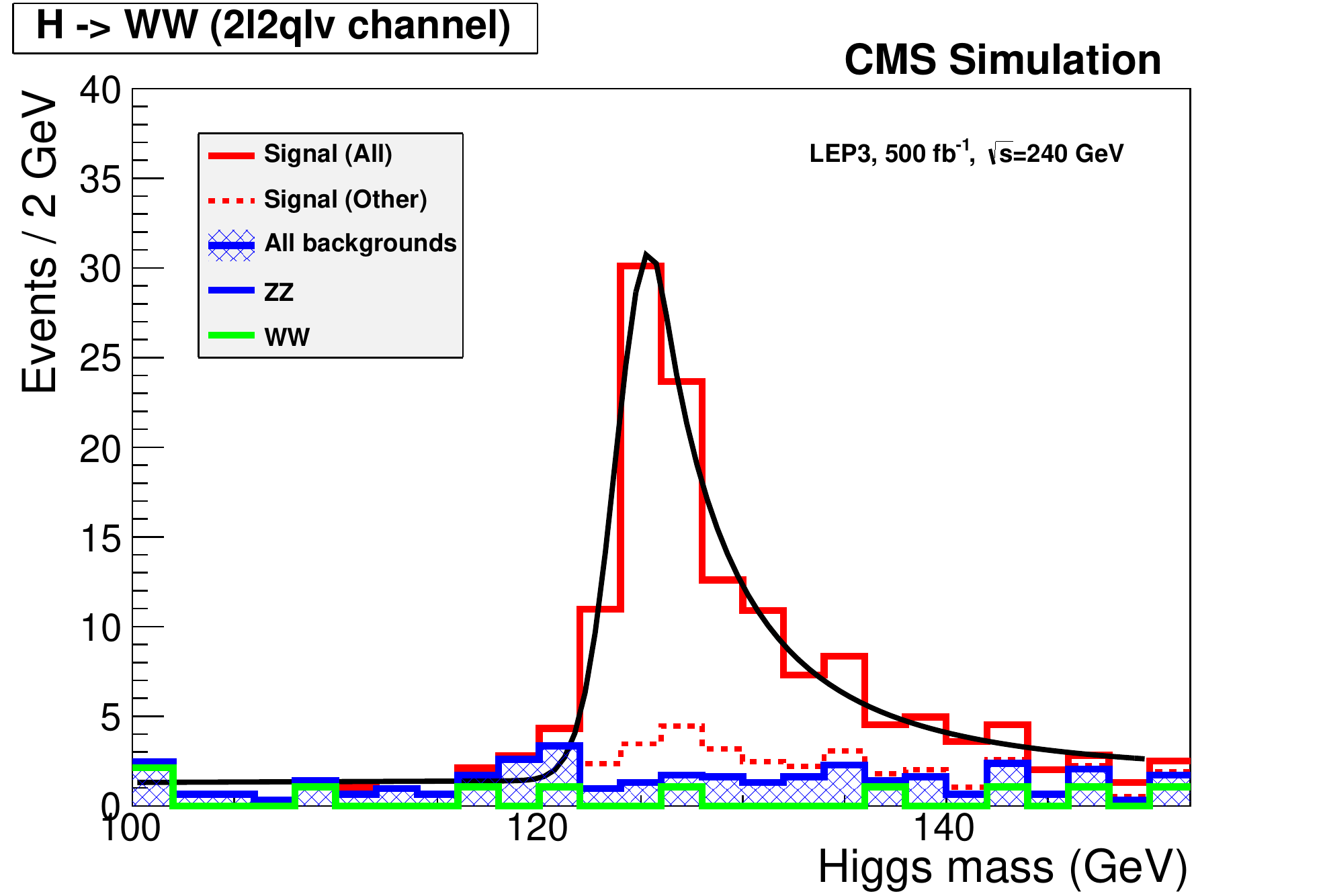}
\caption{\small Distribution of the reconstructed Higgs boson mass (labelled ``Higgs mass'') in the ${\rm ZH} \to \ell^+\ell^- {\rm W}^+{\rm W}^-$ channel, for the signal (hollow histogram) and all backgrounds (shaded histogram), in the fully hadronic (left) and semi-leptonic (right) decays of the W pair. The dashed histogram indicates the Higgs boson decays other than WW.}
\label{fig:llww}
\end{center}
\end{figure*}

As this level, the signal in the fully hadronic channel is still dominated by ${\rm H \to b \bar b}$ decays. After an anti-b-tagging cut is applied. the signal contamination of other Higgs decays is reduced to 50\% (22\% ${\rm H \to b \bar b}$, 6\% ${\rm H \to ZZ}$ and 22\% ${\rm H \to c \bar c\ or\ gg}$). In the semi-leptonic channel, the signal contamination amounts to 10\% and is largely dominated by ${\rm H \to b \bar b}$ decays. The distribution of the mass recoiling against the lepton pair is shown in Fig.~\ref{fig:llww} for the fully hadronic (left) and the semi-leptonic (right) decays of the W pair. The signal is fit to a Crystal Ball function and the backgrounds to a first order polynomial. From the estimate of the accuracy for $\sigma_{\rm HZ} \times {\rm BR(H}\to{\rm b\bar b})$ presented in Section~\ref{sec:bb}, a precision of 11.7\% on $\sigma_{\rm HZ} \times {\rm BR(H}\to {\rm W^+W^-})$ is obtained in the semi-leptonic channel. For the fully hadronic channel, the uncertainty on $\sigma_{\rm HZ} \times {\rm BR(H}\to{\rm c\bar c\ or\ gg})$ estimated in Section~\ref{sec:bb} is also needed, and an uncertainty of 8\% on the ratio ${\rm BR(H}\to{\rm ZZ)} / {\rm BR(H}\to{\rm WW)}$ is used, as can be expected from the 14 TeV run at the LHC. With all these estimates, a precision of 12.7\% on $\sigma_{\rm HZ} \times {\rm BR(H}\to {\rm W^+W^-})$ can be achieved in this channel.
 
\subsubsection{Channels with hadronic Z decays}

Events with two oppositely-charged leptons, either of same flavours (SF) or of opposite flavours (OF), expected to come from the fully leptonic decays of the W pair, are selected as indicated in Section~\ref{sec:hz}. The dilepton invariant mass is required to be between 10 and 70\,GeV/$c^2$ and its energy must exceed 80 GeV. To account for the neutrinos arising from the W decays, the total missing transverse momentum is required to be larger than 25 GeV/$c$. The system recoiling to the lepton pair must consist of at least ten charged hadrons and at least two jets, with an invariant mass compatible with the nominal Z mass, within $\pm 25$\,GeV/$c^2$.

\begin{figure*}[hbtp]
\begin{center} 
\includegraphics[width=0.48\textwidth]{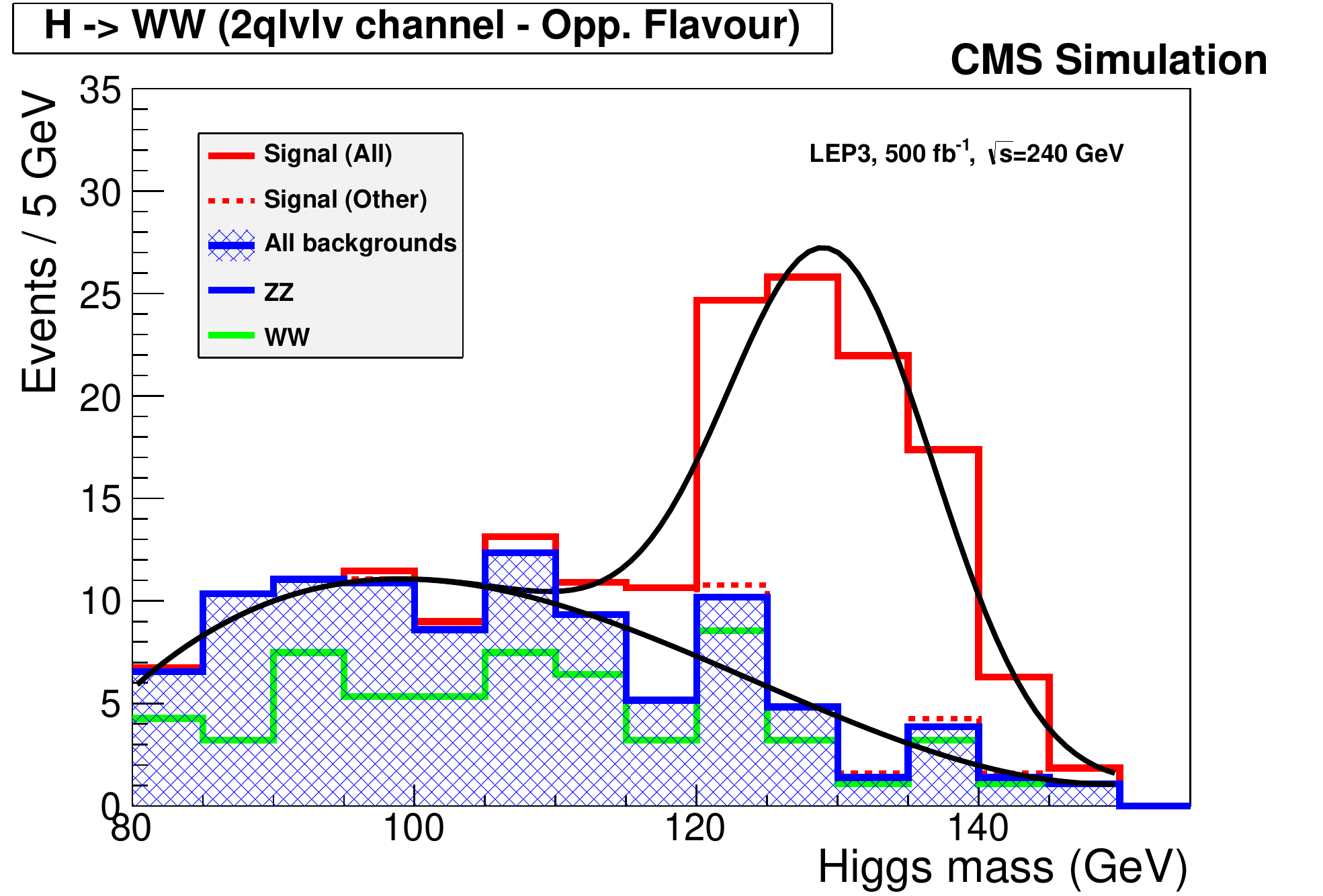}
\includegraphics[width=0.48\textwidth]{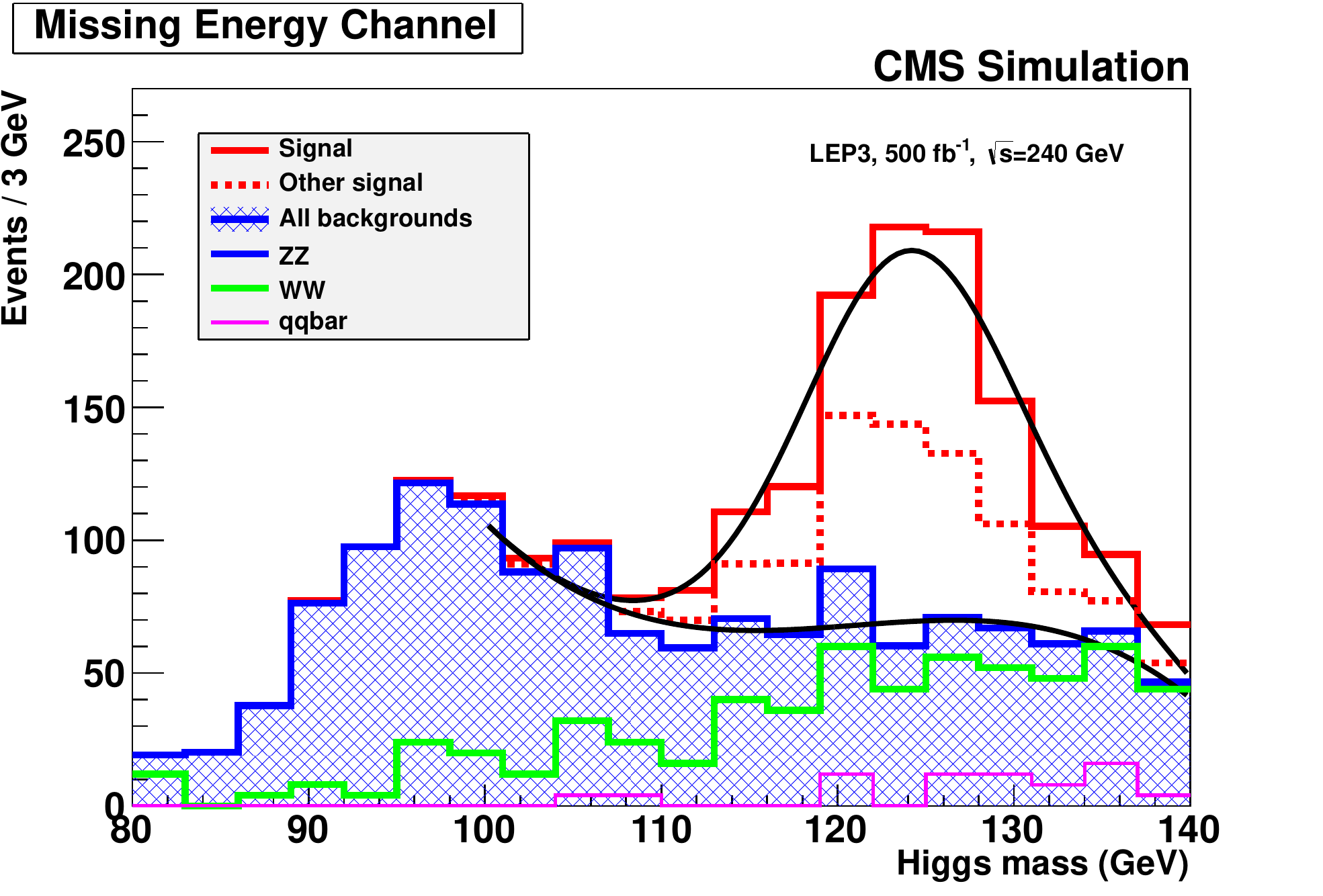}
\caption{\small Distribution of the reconstructed Higgs boson mass (labelled ``Higgs mass'') (left) in the ${\rm ZH} \to {\rm q\bar q} {\rm W}^+{\rm W}^-$ channel, in the fully leptonic decays of the W pair, with  opposite flavours for the two leptons, and (right) in the in the ${\rm ZH} \to \nu\bar\nu {\rm W}^+{\rm W}^-$ channel, with fully hadronic decay of the W pair; for the signal (hollow histogram) and all backgrounds (shaded histogram). The dashed histogram indicates the Higgs boson decays other than WW.}
\label{fig:qqww}
\end{center}
\end{figure*}

The distribution of the mass recoiling against the hadronic system is shown in Fig.~\ref{fig:qqww} (left) for the events with opposite-flavour leptons. (The background in the same-flavour analysis makes this final state statistically less interesting.)  The signal is fit to a Gaussian and the backgrounds to a third-order polynomial. Here, the signal contamination by other Higgs decays is minute (3\%), and is dominated by ${\rm H} \to \tau^+\tau^-$ decays. From the estimate of the accuracy for $\sigma_{\rm HZ} \times {\rm BR(H}\to \tau^+\tau^-)$ presented in Section~\ref{sec:htt}, a precision of 12.8\% on $\sigma_{\rm HZ} \times {\rm BR(H}\to {\rm W^+W^-})$ can be achieved with this hadronic channel. It should be noted that the mass resolution could be substantially improved by rescaling the recoiling hadronic system energy and momentum to constrain its mass to the nominal Z mass.

\subsubsection{Channels with invisible Z decays}

When the Z decays into a pair of neutrinos, the W pair is requested to decay hadronically to avoid the presence of additional missing energy, which would otherwise spoil the Higgs mass determination. Events with at least four jets are therefore selected. Events in which at least one jet consists of an identified electron, an identified muon, an identified tau, or less than three charged hadrons, are rejected, so as to reduce the semileptonic decays of the W pair. To select events with a Z decaying into a neutrino pair, the missing invariant mass is required to exceed 75\,GeV/$c^2$ and the total missing momentum must be larger than 30\,GeV/$c$. To reject hadronic events with a photon radiated by the beam, the angle between the missing momentum and the beam direction is required to be larger than 25 degrees. 

The jets are then reclustered starting from the two jets with the smallest invariant mass, and iterating, until events are formed with only four jets. At least three of the four jets must have an energy larger than 15\,GeV. The jet pair with an invariant mass closest to the W mass is identified, and the invariant mass of the complementary jet pair is required to be larger than 10\,Gev/$c^2$. To reject about half of the Higgs boson decays into ${\rm b \bar b}$, no jet can have a CSV value larger than 0.4. Finally, the jet reclustering proceeds until two jets are reconstructed, and the angle between the two jets is required to exceed 100 degrees, which rejects efficiently a large fraction of the ZZ background.

The jet energies and momenta are rescaled by a unique multiplicative factor aimed at constraining the missing mass to the nominal Z mass. The distribution of the rescaled visible invariant mass is displayed in Fig.~\ref{fig:qqww} (right). The signal is fit to a Gaussian and the background to a third-order polynomial. The signal consists of 60\%  ${\rm H \to WW}$, and the signal contamination composition is similar to that of the sample with fully hadronic WW decays accompanied with a leptonic Z decay (Section \ref{sec:2l4q}). A precision of 9.7\% on $\sigma_{\rm HZ} \times {\rm BR(H}\to {\rm W^+W^-})$ can be achieved.

\subsubsection{Combination of all channels}

Altogether, the combined precision reaches 5.7\%. The most abundant channel, with hadronic decays of the Z and the two Ws, hence leading to a six-jet final state, is under study, and will be added in a future version of this note. If the large WW, ZZ and ${\rm q\bar q }$ background can be reduced to a manageable level, a substantial improvement of the combined accuracy is to be expected. The above study, however, makes is clear that a simultaneous fit of all branching ratios, from the contribution of each decay channel in the different analyses, is needed. This global fit will be implemented in a future version of this note, together with the analysis of additional decay channels (ZZ, ${\rm c \bar c}$, and gg).

\subsection{Measurement of \boldmath{$\sigma_{\rm HZ} \times {\rm BR(H}\to \gamma\gamma)$}}

The branching fraction of the Higgs boson in $\gamma\gamma$ is minute, with just above 250 such events are expected within 500\,${\rm fb}^{-1}$. It is enough, however, to enable the corresponding cross-section measurement. The main background consists of events with double radiative return to the Z mass, ${\rm e^+e^- \to Z} \gamma \gamma$, with the two photons in the detector acceptance. Events with at least two photons of energy larger than 40 GeV in the tracker acceptance are selected. If several such pairs exist in a given event, that with the mass recoiling against the two photons closest to the nominal Z mass is chosen as the "Higgs candidate". The isolation of a photon is defined as the energy sum of the particles reconstructed in a cone around the photon direction divided by the photon energy. The sum of the two photon isolations must be smaller than 0.4, to reject hadronic events with two energetic $\pi^0$s. About 85\% of the ${\rm H} \to \gamma\gamma$ events are selected at this level.

As radiative returns to the Z mass tend to preferentially produce photons close to the beam directions, a large photon-photon invariant mass can be obtained when one photon is emitted backward and the other is emitted forward, as opposed to the signal where photons tend to be emitted centrally. To reject a large fraction of this background, it is required that the pseudo-rapidity gap between the two photons be smaller than 1.8. It is also required that the direction of the Higgs candidate momentum make an angle larger than 25 degrees with respect to the beam axis.

The efficiency of this selection is almost 60\%. The di-photon invariant mass distribution is shown in the graph of Fig.~\ref{fig:hgaga} (left). The background is fit to a third-order polynomial from the side bands, and the signal to a Gaussian. The $\sigma_{\rm HZ} \times {\rm BR(H}\to \gamma\gamma)$ value can be determined with a 14\% accuracy. Substantial improvement, conservatively not included here, is expected to come from the use of the photon energy regression algorithm, currently used in the ${\rm H}\to \gamma\gamma$ analysis at LHC.

\subsection{Measurement of \boldmath{$\sigma_{\rm HZ} \times {\rm BR(H}\to \mu^+\mu^-)$}}

With 22 events expected in an integrated luminosity of 500\,${\rm fb}^{-1}$, it is difficult to achieve any meaningful measurement with only one experiment. The possibility of having four detectors at LEP3 is a clear advantage here. The study is therefore done with the four detectors for this channel, {\it i.e.}, with an equivalent integrated luminosity of 2\,${\rm ab}^{-1}$, in which about 90 ${\rm H} \to \mu^+\mu^-$ events are expected. The four detectors are conservatively assumed to have the same performance as the current CMS detector. 

The event selection proceeds as follows. First, two oppositely charged muons are requested with a relative isolation, determined from all the particles reconstructed in a small cone around the muon momentum direction, smaller than 0.2. This first criterion in 90\% efficient on the signal, and rejects effectively the bulk of the ${\rm q\bar q}$, W(e)$\nu$ events, $\tau^+\tau^-$, and Bhabha events. The mass recoiling to the system formed by the two muons and the recovered bremsstrahlung photons (called the Higgs candidate) is then required to be compatible with the Z mass, i.e., between 80 and 110\,GeV/$c^2$. To reject the large WW background with both Ws decaying to $\mu\nu$, the Higgs candidate is required to be accompanied by two visible "jets". This cut efficiently also rejects the even larger $\mu^+\mu^- (\gamma)$ background, when the photon escapes along the beam axis. It rejects 20\% of the signal as well, when the Z decays into a pair of neutrinos. To reject $\mu^+\mu^-\gamma\gamma$ events where the two photons are in the detector acceptance, the electromagnetic fraction of at least one of the two jets must be smaller than 80\%. This cut also rejects the large Ze(e) background, as well as 3.4\% of the signal, when the Z decays into a pair of electrons. 

\begin{figure*}[hbtp]
\begin{center} 
\includegraphics[width=0.48\textwidth]{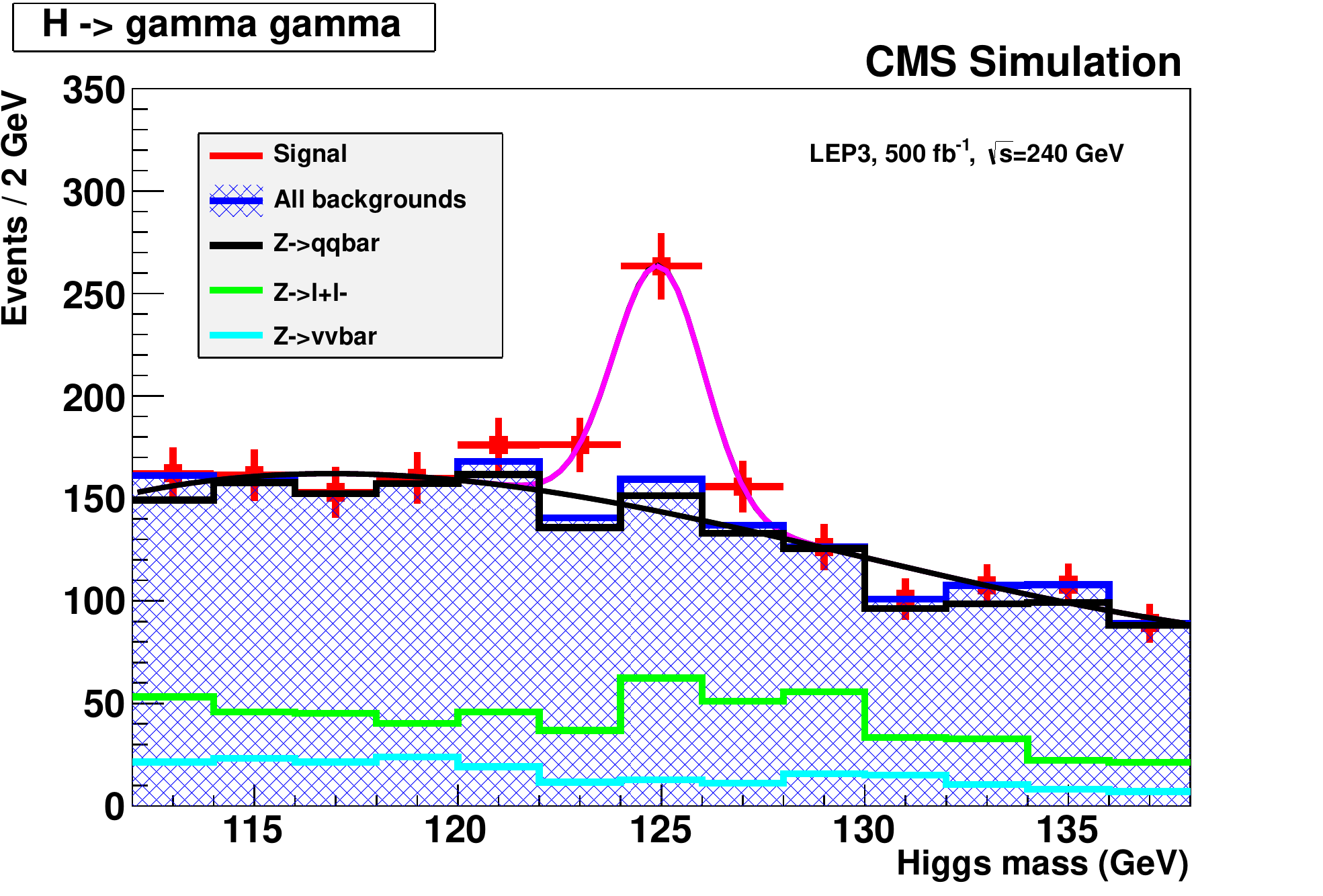}
\includegraphics[width=0.48\textwidth]{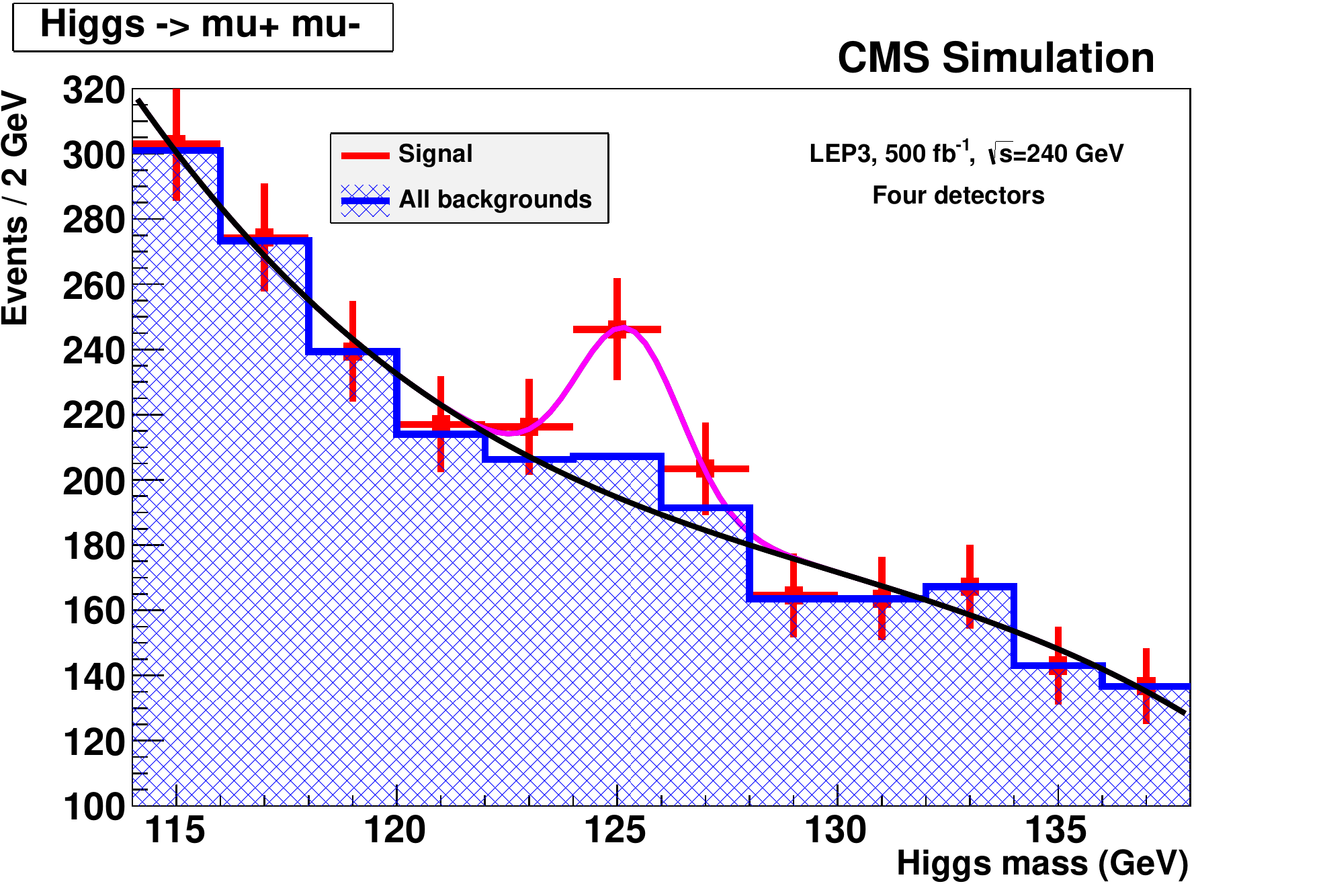}
\caption{\small Distribution of the di-photon invariant mass (labelled ``Higgs mass'') in the  ${\rm H} \to \gamma\gamma$ channel (left) and in the ${\rm H} \to \mu^+\mu^-$ channel (right), for the HZ signal (hollow histogram) and all backgrounds (shaded histogram).}
\label{fig:hgaga}
\end{center}
\end{figure*}

The di-muon invariant mass distribution is shown in Fig.~\ref{fig:hgaga} (right), for the statistics accumulated by four experiments. The background, mostly originating from ZZ production, is fit from the side bands to a third-order polynomial, and the signal is fit to a Gaussian. A $4\sigma$ excess is well visible and the $\sigma_{\rm HZ} \times {\rm BR(H} \to \mu^+\mu^-)$ value can be measured with a precision of 28\%. This performance is already enough to ascertain the Higgs decay into a pair of muons in view of a future muon collider project. It is also likely, however, that the trackers of both ATLAS and CMS would be refurbished to lead to a better muon momentum resolution if they were to operate with LEP3. The two new detectors would be tuned with this performance in mind, thus leading to a much better dimuon mass resolution, hence a much better precision on the $\mu^+\mu^-$ branching fraction.

\subsection{Measurement of the Higgs boson mass}

The fits presented in the previous sections all return an estimate for the Higgs boson mass. The statistical accuracies of these estimates are compiled in Table~\ref{tab:mass}.

\begin{table*}[htbH]
\begin{center}
\topcaption{\small The statistical precision on the Higgs boson mass in some of the channels studied in this Section, for an integrated luminosity of 500\,${\rm fb}^{-1}$.
\label{tab:mass}}
\begin{tabular}{|l|r|}
\hline\hline 
Final state & Accuracy (MeV/$c^2$) \\
\hline\hline
$\ell^+\ell^-$ H & 80 \\
\hline
${\rm q \bar q b \bar b}$ & 109 \\
\hline
$\nu\bar\nu{\rm b\bar b}$ & 154 \\
\hline
${\rm q \bar q} \tau^+\tau^-$ & 225 \\
\hline
$\nu\bar\nu{\rm W^+W^-}$ & 810 \\
\hline
${\rm H}\to \gamma\gamma$ & 160 \\
\hline
${\rm H}\to \mu^+\mu^-$ & 580 \\
\hline\hline
\end{tabular}
\end{center}
\end{table*}
The combined statistical precision amounts to 53\,MeV/$c^2$. A small bias, originating from the fitting methods and functions, is observed ($\sim 200$\,MeV/$c^2$). It can easily be corrected for with well established methods, used for the W mass measurement, for example.

\section{Conclusion: Comparison with the ILC physics potential}
\label{sec:LC}

As already stated, the present work is not as comprehensive and refined as the ILC studies, carried out and optimized for the past 20 years. In addition, and unlike the proposed design for the ILC detector, the CMS detector and software were neither thought nor optimized for ${\rm e^+e^-}$ collisions. It is nevertheless instructive to compare the expected performance of the very conservative approach adopted here with the optimized performance of a linear collider (taken from Ref.~\cite{ilc}), at least for the channels studied in this note. This comparison is shown in Table~\ref{tab:comp} for two LEP3 configurations, {\it (i)} with CMS and ATLAS only, under the reasonable assumption that the ATLAS performance are similar to those of CMS when run in ${\rm e^+e^-}$ collisions; and {\it (ii)} with two additional detectors with performance similar to those of CMS. 

\begin{table*}[htbH]
\begin{center}
\topcaption{\small The precision (or 95\% C.L. sensitivity for the invisible decay) on the Higgs boson cross sections and couplings obtained from studies of the Higgsstrahlung process, with five years of running at the ILC, at LEP3 with CMS and ATLAS, and at LEP3 with two additional detectors of similar performance. The numbers for the ILC were obtained with $m_{\rm H}=120$\,GeV/$c^2$, $\sqrt{s}=250$\,GeV, and leading-order cross sections, while those for LEP3 were conservatively obtained with $m_{\rm H}=125$\,GeV/$c^2$, $\sqrt{s}=240$\,GeV, and next-to-next-to-leading-order cross section. The number from TLEP assume two detectors with performance similar to those developed for ILC. The LEP3 missing couplings will be added in a forthcoming update of this note. For completeness, the LHC sensitivity with 300 and 3000${\rm fb}^{-1}$ at 13\,TeV is also given. For 300\,${\rm fb}^{-1}$, the model-independent projections from {\tt SFITTER}~\cite{klute} and the simplified-model projection from CMS~\cite{CMSESG} are shown, while only the latter is available HL-LHC. The precision on the Higgs boson mass is indicated in the last row of the table.
\label{tab:comp}}
\begin{tabular}{|l|c|c|c|c|c|c|}
\hline\hline 
  & ILC & LEP3 (2) & LEP3 (4) & TLEP (2) & LHC (300) & HL-LHC \\
\hline\hline
$\sigma_{\rm HZ}$ & 3\% & 1.9\% & 1.3\% & 0.7\% & -- & -- \\
\hline
$\sigma_{\rm HZ} \times {\rm BR(H}\to{\rm b\bar b})$ & 1\% & 0.8\% & 0.5\% & 0.2\% & -- & -- \\
\hline
$\sigma_{\rm HZ} \times {\rm BR(H}\to \tau^+\tau^-)$ & 6\% & 3.0\% & 2.2\% & 1.3\% & -- & -- \\
\hline
$\sigma_{\rm HZ} \times {\rm BR(H}\to W^+W^-)$ & 8\% & 3.6\% & 2.5\% & 1.6\% & -- & -- \\
\hline
$\sigma_{\rm HZ} \times {\rm BR(H}\to \gamma\gamma)$ & ? & 9.5\% & 6.6\% & 4.2\%& -- & -- \\
\hline
$\sigma_{\rm HZ} \times {\rm BR(H}\to \mu^+\mu^-)$ & -- & -- & 28\% & 17\%& -- & -- \\
\hline
$\sigma_{\rm HZ} \times {\rm BR(H}\to {\rm invisible})$ & ? & 1\% & 0.7\% & 0.4\% & -- & -- \\
\hline
$g_{\rm HZZ}$ & 1.5\% & 0.9\% & 0.6\% & 0.3\% & 13\%/5.7\% & 4.5\%\\
\hline
$g_{\rm Hbb}$ & 1.6\% & 1.0\% & 0.7\% & 0.4\% & 21\%/14.5\% & 11\% \\
\hline
$g_{{\rm H}\tau\tau}$ & 3\% & 2.0\% & 1.5\% & 0.6\% & 13\%/8.5\% & 5.4\% \\
\hline
$g_{\rm Hcc}$ & 4\% & ? & ? & 0.9\% & ?/? & ? \\
\hline
$g_{\rm HWW}$ & 4\% & 2.2\% & 1.5\% & 0.9\% & 11\%/5.7\% & 4.5\% \\
\hline
$g_{{\rm H}\gamma\gamma}$ & ? & 4.9\% & 3.4\% & 2.2\% & ?/6.5\% & 5.4\% \\
\hline
$g_{{\rm H}\mu\mu}$ & -- & -- & 14\% & 9\% & ? & ? \\
\hline
$g_{\rm Htt}$ & -- & -- & -- & -- & 14\% & 8\% \\
\hline
$m_{\rm H}$ (MeV/$c^2$) & 50 & 37 & 26 & 11 & 100 & 100 \\
\hline\hline
\end{tabular}
\end{center}
\end{table*}
Conservatively, none of the obvious analysis improvements and none of the planned detector/software improvements mentioned in the previous sections (which could reduce some of the above figures by factors larger than 2) are included in the LEP3 estimates. The assumption that the two additional detectors would have performance similar to those of CMS is also very conservative, as the additional detectors would most likely resemble those currently proposed for ILC, optimized for ${\rm e^+e^-}$ collisions and for particle-flow reconstruction. It must also be mentioned that the simulations presented in this note are extremely realistic, as they are based on the full simulation of an existing/operational detector, including all its noisy and dead cells, and the description of the detector material tuned with real data. In spite of this conservative approach, the expected precision on the Higgs boson couplings studied so far is significantly better at LEP3, typically a factor 2 to 3, than the ILC most recent projections. The TLEP project promises even better precision when run at 240\,GeV, typically a factor 4 to 5 better than these ILC projections. (When run at 350\,GeV, further improvements are expected for the WW coupling due to the contribution of WW fusion, but these improvements are not included in the Table yet.) The larger instantaneous luminosity and the possibility of operating several detectors with no reduction of this luminosity are decisive advantages of a collider ring over a linear collider in this respect.

The specificities of a collider ring are even more favourable at lower centre-of-mass energies. At the Z peak, for example, instantaneous luminosities more than two orders of magnitude larger than with a linear collider can be achieved. With four detectors, it would turn the linear collider ``Giga-Z factory'' into a ``Tera-Z factory''. In principle, the precision of all Z pole measurements could then be reduced by factors ranging from 25 to 100 with respect to the LEP precision, and by factors up to 10 with respect to the Giga-Z option. Similarly, the precision on the W mass measurement, which can be carried out both at the WW threshold and at the Higgs factory, could reach a fraction of MeV/$c^2$, a factor 50 smaller than the current precision. The top quark mass could be measured with a precision of 50\,MeV/$c^2$ or better with TLEP.

A lot of practical issues are of course to be studied in detail, but the large luminosity, the number of detectors, the relaxed beamstrahlung conditions, the negligible pileup rate, or the unique beam energy measurement, all make a collider ring the ideal place to be towards precision Higgs physics in ${\rm e^+e^-}$ collisions during the next decade.

The ILC physics programme, however, includes expensive phases at centre-of-mass energies of 500\,GeV and 1 TeV, which a collider ring cannot reach. If the 13 TeV run of the LHC reveals no new physics to be studied with precision at ILC, the main purpose of these phases would be the important measurements of the Higgs coupling to the top quark, via ${\rm e^+e^- \to t\bar t H}$ production at 500\,GeV, and of the Higgs trilinear coupling, via double Higgs boson production through WW fusion, ${\rm e^+e^-} \to \nu_{\rm e}\bar\nu_{\rm e} {\rm H H}$. Preliminary studies indicate that a precision of 10\% and 20\% on these couplings is achievable at 500\,GeV and 1\,TeV~, respectively\cite{ilc}. This precision is comparable to recent projections with HL-LHC from the ATLAS collaboration~\cite{ATLASESG}, and even better prospects can be inferred with HE-LHC. The really unique physics programme of an ${\rm e^+e^-}$ collider therefore seems to be at lower centre-of-mass energies.

From these considerations, the high-energy-physics programme for CERN, with the ATLAS and CMS experiments, could be imagined as follows in the next 30 years: {\it (i)} the approved LHC run until 2022, to collect up 300\,${\rm fb}^{-1}$ of pp collisions at $\sqrt{s} = 13$\,TeV in ATLAS and in CMS; {\it (ii)} the LEP3 physics programme at the Z pole, the WW threshold and the ZH threshold, between 2025 and 2032, with ATLAS and CMS, and possibly two additional detectors optimized for ${\rm e^+e^-}$ collisions; and {\it (iii)} a high-energy pp collider, HE-LHC, to collect data at $\sqrt{s} = 33$\,TeV in refurbished versions ATLAS and in CMS, from 2035 onwards.

\section*{Acknowledgements}
We are indebted to Alain Blondel for enlightening discussions about the LEP3 physics, and to Gigi Rolandi for his careful reading and commenting of the manuscript. We would also like to thank Frank Zimmermann and Mike Koratzinos for sharing their deep knowledge of the machine aspects with us.

\bibliography{auto_generated}   

\end{document}